\documentstyle[12pt,epsf]{article}

\catcode`\@=11
\@addtoreset{equation}{section}

\def\l#1{{\lambda}_{#1}}
\def\lp#1{{\lambda^{\prime}}_{#1}}
\def\tlp#1{{\tilde{\lambda^{\prime}}_{#1}}}
\def\lpp#1{{\lambda^{\prime\prime}_{#1}}}
\def\C#1{{C}_{#1}}
\def\Cp#1{{C^{\prime}_{#1}}}
\def\Cpp#1{{C^{\prime\prime}_{#1}}}

\def\lesssim{\stackrel{<}{\sim}}

\def\be{\begin{equation}}
\def\ee{\end{equation}}
\def\br{\begin{eqnarray}}
\def\er{\end{eqnarray}}

\def\bsg{b \rightarrow s \gamma}

\begin{document}

\baselineskip 24pt

\begin{titlepage}
\begin{flushright}
SUSX-TH/96-009, OUTP-96-34P, IEM-FT-135/96\\
hep-ph/9609443
\end{flushright}
\vspace{.2in}
\begin{center}

{\large{\bf $R$-Parity Violation and
Quark Flavour Violation}}
\bigskip \\
{\large B.~de Carlos${}^1$ and P.~L.~White${}^2$}\\ 
\vskip 0.2in
{\it 
${}^1$ Centre for Theoretical Physics, 
University of Sussex, \\ Falmer, Brighton BN1 9QH, UK. \\
Email: {\tt B.De-Carlos@sussex.ac.uk} \\}
\vskip 0.2in
{\it
${}^2$ Theoretical Physics, University of Oxford, \\
1 Keble Road, Oxford OX1 3NP, UK. \\
Email: {\tt plw@thphys.ox.ac.uk}
}
\\ \vspace{.5in}
{\bf Abstract} \smallskip \end{center} \setcounter{page}{0}
We investigate the implications of $R$-Parity violation (RPV) for 
quark flavour violation both by constraining the sneutrino masses to
be positive and by studying the processes $b\to s\gamma$ and
$K^0-\bar K^0$ mixing. In the latter there are two major
contributions, one from ``direct'' one loop diagrams involving RPV 
couplings, and one from the ``indirect'' contributions generated by 
the renormalisation group. We compare the effects and discuss the 
implications of our results.

\end{titlepage}

\section{Introduction}
One of the most promising candidates for physics beyond the so-called 
Standard Model (SM) is that of supersymmetry (SUSY) \cite{revs}. SUSY
has the highly attractive properties of giving a natural explanation
to the hierarchy problem of how it is possible to have a low energy 
theory containing light scalars (the Higgs) when the ultimate theory 
must include states with masses of order the Planck mass. This allows 
the ultimate hope of constructing a theory of gauge unification and 
fermion (and superpartner) masses defined at a scale near the Planck 
mass yet whose structure and parameters may be deduced from physics at
accessible scales.

In this paper we shall be concerned with the implications of a
particular possible feature of SUSY, namely that of $R$-Parity
violation (RPV) \cite{rpv,barbm,suzuki}. $R$-Parity is a $Z\!\!\! Z_2$
symmetry of both the SM and its minimal SUSY extension, the MSSM,
under which all of the SM particles have charge 0, while all their
SUSY partners have charge 1. Its implications include the stability of
the lightest supersymmetric particle (LSP), and hence the typical SUSY
collider signatures of missing $E_T$ and the existence of a source of 
dark matter. Its violation changes both the implied cosmology and the
expected collider signatures, allowing such effects as LSPs decaying
inside the detector and leptoquarks \cite{bhat}. In addition to these,
further constraints on RPV can be derived by considering experimental
limits on rare decays \cite{bgh,bpw,bgnn}, and by the demands of
proton stability. In practice it is usual to evade the problems of 
proton decay by considering either baryon number or lepton number 
violation but not both simultaneously.

$R$-Parity is violated by the superpotential and soft potential
\br
W&=&
   h^u_{ij}Q_iH_2u_j + h^d_{ij} Q_iH_1d_j + h^e_{ij} L_iH_1e_j \cr 
   &&+ \frac{1}{2}\l{ijk}L_iL_je_k + \lp{ijk}L_iQ_jd_k
                + \frac{1}{2}\lpp{ijk}u_id_jd_k \cr
   &&+ \mu_4 H_1H_2 + \mu_iL_iH_2 \cr
V_{\rm soft}&=&
   \eta^u_{ij}Q_iH_2u_j + \eta^d_{ij} Q_iH_1d_j
 + \eta^e_{ij}L_iH_1e_j + {\rm h.c.} \cr
   &&+ \frac{1}{2}\C{ijk}L_iL_je_k + \Cp{ijk}L_iQ_jd_k
                + \frac{1}{2}\Cpp{ijk}u_id_jd_k + {\rm h.c.}  \cr
   &&+ \frac{1}{2}M_a\lambda_a^c\lambda_a
                + \sum_{a,b} m_{ab}^2 \varphi_a\bar\varphi_b \cr
   &&+ D_4 H_1H_2 + D_iL_iH_2 + {\rm h.c.}
\er
From the point of view of deriving constraints on the $R$-Parity
violating couplings in the model, the most extensively studied
couplings are the dimensionless couplings $\l{}$, $\lp{}$, and
$\lpp{}$, which directly generate many effects which can be
experimentally limited. The extra soft terms by definition mostly
couple only heavy SUSY particles and hence are relevant mostly because
of their impact on the Renormalisation Group Equations (RGEs),
although they can have significant effects on the neutrino-neutralino 
and Higgs boson-sneutrino sectors \cite{suzuki,rv,paper1}.

In SUSY models, flavour-changing effects may be caused by the
existence of off-diagonal terms in the sfermion mass matrices in the 
basis in which the fermion masses are diagonal, in which there are no 
tree level SM flavour-changing neutral currents (FCNC). Such
flavour-violating soft masses can be generated either from the high
energy theory such a GUT directly, or else through the RGEs by
couplings which violate flavour symmetries, such as Yukawa couplings
mediated by the CKM matrix and both possibilities have been studied
extensively \cite{hagelin,otherfcnc}. However, the inclusion of the
RPV couplings in the RGEs allows many extra violations of quark and 
lepton flavour.

In a previous analysis \cite{paper1}, we presented the renormalisation
group equations (RGEs) for the couplings of the full $R$-Parity
violating sector of the model, and investigated the implications of
typical scenarios at the GUT scale for the generation of neutrino
masses and the decay $\mu\to e\gamma$. Contributions can conveniently
be split into two categories which we term ``direct'' (where the
flavour violation occurs directly through $R$-Parity violating
vertices in the diagrams) and ``indirect'' (where $R$-Parity violation
induces flavour violation through the RGEs on the soft masses). We
found that the indirect effects were typically large, often orders of 
magnitude larger than the direct ones. However, the extremely
complicated dependence on the spectrum and the existence of many 
possible cancellations in the amplitude render the process of deriving
bounds impossible except on an order of magnitude basis.

The intention of this paper is to extend the work of reference
\cite{paper1} to consider quark flavour violation (QFV) effects (for
recent work on the subject see ref.~\cite{rel}) .
Following this introduction, we give a short discussion on a new bound
from requiring the sneutrino masses to be above their experimental
limits in section 2, while section 3 is devoted to the effects of RPV
on $b\to s\gamma$, and section 4 to the $K^0-\bar K^0$ mixing term
$\Delta m_K$. Many useful formulae for our RGEs and definitions are
relegated to the appendices, while section 5 contains our conclusions.

\section{Bounds from Sneutrino Masses}
We now comment on a new simple bound on certain of the Yukawa
couplings, which is associated with our assumptions of unification of
couplings at a high scale. If we add a new Yukawa coupling associated
with the sneutrinos (here $\lp{}$), the mass of the sneutrino is
driven down by its effect on the RGEs, and so increasing the value of 
the coupling will ultimately drive the mass of the sneutrino below the
experimental limit. The low energy values of the lepton masses and the
corresponding soft masses are given by
\br
m^2_{L_i}&=& m_0^2 + 0.51 M_{1/2}^2
 - \sum_{jk}\lp{ijk}^2(M_{GUT})
 \bigl [ 13m_0^2 + 49M_{1/2}^2 - 1.5 M_{1/2}A_0 - 12 A_0^2\bigr ]
\cr
m^2_{\tilde\nu_i}&=& m^2_{L_i} + \frac{1}{2}M_Z^2\cos2\beta
\er
where we have assumed universal masses at the unification scale and
solved the RGEs numerically.
The experimental limit on sneutrinos is 37 GeV \cite{pdb} (assuming
that the sneutrinos are not degenerate), so that we conclude that the
Yukawa couplings $\lp{}$ must be bounded by
\be
\sum_{jk}\lp{ijk}^2(M_{GUT}) < \frac
{m_0^2 + 0.51 M_{1/2}^2 + \frac{1}{2}M_Z^2\cos2\beta - (37\hbox{GeV})^2}
{13m_0^2 + 49M_{1/2}^2 - 1.5 M_{1/2}A_0 - 12 A_0^2}
\label{sneutmassbound}
\ee
Given that masses for squarks and sleptons as low as 100 GeV are barely
tenable in a unified framework with present collider limits, this
constraint is the tightest available on certain $\lp{}$. For example,
if we set $A_0\simeq 0$, $M_{1/2}=m_0=200$ GeV and $\tan\beta=10$ then
this gives a bound on all of the $\lp{}$ of 0.15, while if two of the
$\lp{}$ are equal then they must be less than 0.11, bounds which
become only very slightly weaker with increasing soft masses, while
the corresponding numbers with $m_0$ and $M_{1/2}$ both 100 GeV are
0.12 and 0.09. Note that these couplings are at the GUT scale, and the
electroweak scale values are then around three times larger, while
there is an error of at least 10 to 20\% from the uncertainty in the
value of the strong coupling.

Although we have neglected the effect of the tau Yukawa coupling, this
in fact only makes the bound for the third generation rather tighter
by reducing the sneutrino mass for that generation still further. We 
can also derive bounds on the $\l{}$ and $\lpp{}$ couplings from
similar arguments, but the results are not very restrictive.

\section{$b\to s\gamma$}
\subsection{Contributions}
The rare decay process $b\to s\gamma$ has a branching rate which can 
be deduced from the decay $B\to K^*\gamma$ which has been measured
\cite{CLEO} to obtain a value
\be
\hbox{B}(b\to s\gamma)=(2.32\pm 0.57 \pm 0.35) \times 10^{-4}
\ee
consistent with the standard model result, and hence a 95\% confidence
level limit of
\be
1.0\times 10^{-4} < \hbox{B}(b\to s\gamma) < 4.2 \times 10^{-4}
\label{bsglimit}
\ee
This gives a very strong test of new physics such as supersymmetry,
since SUSY models with light spectra can give large contributions to
this process through charged Higgs and chargino diagrams
\cite{bsgsusy}.

The total branching ratio for the process $\bsg$ can be written (in
units of the BR for the semileptonic $b$ decay) as:
\be
\frac{\hbox{B}(\bsg)}{\hbox{B}(b \rightarrow c e \bar{\nu})} = 
\frac{3 \pi \alpha}{G_F^2 |K_{cb}|^2 I(z)} \left(|\tilde{A}_{LR}|^2 + 
|\tilde{A}_{RL}|^2 \right) F
\label{br}
\ee
where $G_F$ is the Fermi constant, $z=m_c/m_b$, $I(z) = 
1- 8z^2+ 8z^6- 24z^4\log(z)$ is the phase space factor, $K_{ij}$ will
be the different CKM matrix elements, $\tilde{A}_{LR}$,
$\tilde{A}_{RL}$ are the total amplitudes to the LR and RL
transitions respectively, and 
\be
F \sim \left( 1 - \frac{8}{3} \frac{\alpha_s(m_b)}{\pi} \right)
\frac{1}{\kappa(z)}
\ee
contains NLO effects ($\kappa(z)$ being the NLO correction to the
semileptonic decay). Here
\be
\tilde{A}_i = \eta^{16/23} \tilde{A}_i^{\gamma} + \frac{8}{3} 
(\eta^{14/23}-\eta^{16/23}) \tilde{A}_i^{g} + C \tilde{A}_i^0 
\ee
with $i=LR,RL$, $\eta=\alpha_s(M_W)/\alpha_s(m_b)$ and the different
terms are as follows: $\tilde A_{LR,RL}^{\gamma,g}$ are the
coefficients of the effective operators for the $bs\gamma$ and $bsg$
interactions
\br
C^{\gamma}_{LR,RL}&=&
 \frac{e}{4\pi}m_b(\bar s\sigma^{\mu\nu}P_{R,L}b)F_{\mu\nu} \cr
C^g_{LR,RL}&=&
 \frac{g_3}{4\pi}m_b(\bar s^i\sigma^{\mu\nu}P_{R,L}b_j)
           G^a_{\mu\nu}T^a_i{}^j
\er
$\tilde A_i^0=-\alpha_W K^*_{ts} K_{tb}/M_W^2$ is from the coefficient
of an operator
\be
C^0_i= [\bar sP_Lc] [\bar cP_Lb] \;\;,
\ee
and $C$ stands for the leading logarithmic QCD corrections (for a
complete list of references see \cite{bsgQCD}).

The LR amplitudes can be divided into an $R$-Parity conserving part 
plus an RPV one; the former has been calculated in ref.~\cite{bert}, 
and corresponds mainly to the contributions coming from the SM diagram
plus those with top quark and charged Higgs, and stops/scharms and 
charginos running in the loop, plus smaller contributions from loops 
with neutralinos or gluinos and d--type squarks, which are generated 
due to QFV explicitly through the CKM matrices. Their expressions are
\br
\tilde{A}^{\gamma,g}_{SM} & = & 
\frac{\alpha_W}{2}
K^*_{ts} K_{tb} \frac{3}{M_W^2} \frac{m_t^2}{M_W^2}
f_{\gamma,g}^{(1)} \left( \frac{m_t^2}{M_W^2} \right) \\
\tilde{A}^{\gamma,g}_{H^-} & = &
\frac{\alpha_W}{2} K^*_{ts} K_{tb}
\frac{1}{M_W^2} \frac{m_t^2}{m_{H^-}^2} \left[
\frac{1}{\tan^{2} \beta} f_{\gamma,g}^{(1)} \left(
\frac{m_t^2}{m_{H^-}^2} \right)
+ f_{\gamma,g}^{(2)} \left( \frac{m_t^2}{m_{H^-}^2} \right) \right]
 \\
\tilde{A}^{\gamma,g}_{\chi^{-}} & = & - \alpha_W K^*_{ts} K_{tb} 
\sum_{j=1}^{2} \left\{
\frac{-1}{m^2_{\tilde{c}_1}} |V_{j1}|^{2} f_{\gamma,g}^{(3)}
\left( \frac{M^2_{\chi_j^-}}{m_{\tilde{c}_1}^2} \right) \right.
\cr
&& \qquad\qquad + \sum_{k=1}^{2} \frac{1}{m^2_{\tilde{t}_k}}
\left| V_{j1} T_{k1} - \frac{m_t V_{j2} T_{k2}}{\sqrt{2} M_W \sin\beta}
\right|^{2} f_{\gamma,g}^{(3)} 
\left( \frac{M^2_{\chi_j^-}}{m^2_{\tilde{t}_k}} \right) \cr
&& 
- \frac{U_{j2}}{\sqrt{2}\cos\beta} \frac{M_{\chi_j^-}}{M_W}  
 \left[
\frac{-1}{m^2_{\tilde{c}_1}} V_{j1} f_{\gamma,g}^{(4)} \left( 
\frac{M_{\chi_j^-}^2}{m^2_{\tilde{c}_1}} \right) \right. \cr
&&
\qquad \left. \left.
+ \sum_{k=1}^{2} \frac{1}{m^2_{\tilde{t}_k}}
\left( V_{j1} T_{k1} - \frac{m_tV_{j2} T_{k2}}{\sqrt{2}M_W\sin\beta} 
\right) T_{k1} f_{\gamma,g}^{(4)} \left(
\frac{M_{\chi_j^-}^2}{m^2_{\tilde{t}_k}} \right) \right] \right\} \;\; ,
\label{amps}
\er
where our notation is as in ref.~\cite{paper1}. $T$, $B$ are the
orthogonal matrices that diagonalise the stop and sbottom mass
matrices respectively through $T M^2_{\tilde{t}-weak} T^{\dagger} =
M^2_{\tilde{t}-diag}$
where $T_{11}=T_{22}=\cos\theta_t$, $T_{12}=-T_{21}=\sin\theta_t$,
so that we may write the mass eigenstates $|\tilde t^{(1)}>$ and
$|\tilde t^{(2)}>$ as $\cos\theta_t|\tilde t_L>+\sin\theta_t|\tilde t_R>$
and $-\sin\theta_t|\tilde t_L>+\cos\theta_t|\tilde t_R>$ respectively,
and similarly for other flavours. For the first and second generations
with small left-right mixing we shall take $\cos\theta_i=1$, so that
for example $\tilde d^{(1)}_i=\tilde d_L^i$ and $\tilde
d^{(2)}_i=\tilde c_R^i$. For simplicity, we shall use the notation
$\tilde s_1$ rather than $\tilde d_2^{(1)}$, except where confusion
might arise. The different functions $f_{\gamma,g}^{(i)}$ are defined
in Appendix C.

We generate new contributions to both the LR and RL amplitudes from
$R$-parity violating couplings ($\lp{}$ and $\lpp{}$) directly and also
indirectly from the induced QFV soft  terms. Therefore we can write
\be 
\tilde{A}_{LR}^{RPV} = \tilde{A}_{LR}^{\lp{}} + \tilde{A}_{LR}^{\Delta
m_{\chi^-}} + \tilde{A}_{LR}^{\Delta m_{\chi^0}} +
\tilde{A}_{LR}^{\Delta m_{\tilde{g}}} \;\; .
\ee
The direct amplitude is given by
\be
\tilde{A}_{LR}^{\gamma\lp{}}  = Q_d \tilde{A}_{LR}^{g\lp{}} =  
- Q_d \sum_{i,j=1}^{3} \frac{\tlp{i2j} \tlp{i3j}}{4\pi} \left( 
\frac{1}{12} \left[
\frac{\sin^2 \theta_{d_j}}{m^2_{\tilde{d}_j^{(1)}}} + 
\frac{\cos^2 \theta_{d_j}}{m^2_{\tilde{d}_j^{(2)}}} \right] 
 - \frac{1}{m^2_{\tilde{\nu_i}}} F_1(x_{ji}) \right) \; ,
\label{alplr}
\ee
where $\tlp{}$ are the different RPV couplings in the fermion mass
eigenstate basis which, in this case, is related to the weak one by:
\be
\tlp{ijk} = \lp{imk} K_{mj} \; ,
\ee
where a sum over $m$ is understood. Note that this still leaves the 
possibility of generating effects even with only one non-zero RPV 
coupling in the weak basis \cite{ag}. 
Here $x_{ji}={m^2_{d_j}}/{m^2_{\tilde{\nu_i}}}$.

The other amplitudes are:
\br
\tilde{A}_{LR}^{\gamma,g\Delta m_{\chi^-}} & = & 
-\alpha_W K^*_{cs} K_{tb} \sum_{n=1}^{2}
\frac{\Delta m^2_{\tilde{c}_1
\tilde{t}_{n}}}{m^2_{\tilde{t}_n}-m^2_{\tilde{c}_1}} 
\sum_{j=1}^{2}  V_{j1} \left\{ \left( V^{*}_{j1} T^{*}_{n1} - 
\frac{m_t V^{*}_{j2}T^{*}_{n2}}{\sqrt{2} M_W \sin\beta} \right)
\right. \\
& \times & \left. \left( 
\frac{f_{\gamma,g}^{(3)}(x^t_{jn})}{m^2_{\tilde{t}_n}} -
\frac{f_{\gamma,g}^{(3)}(x^c_{j1})}{m^2_{\tilde{c}_1}} \right)  - 
\frac{U_{j2}T^{*}_{n1}}{\sqrt{2} \cos \beta}
\frac{M_{\chi_j^-}}{M_W}
\left( \frac{f_{\gamma,g}^{(4)}(x^t_{jn})}{m^2_{\tilde{t}_n}} -
\frac{f_{\gamma,g}^{(4)}(x^c_{j1})}{m^2_{\tilde{c}_1}} \right) 
\right\} \cr
\tilde{A}_{LR}^{\gamma\Delta m_{\chi^0}} & = &
Q_d \tilde{A}_{LR}^{g\Delta m_{\chi^0}} =  
- 2\alpha_W Q_d \sum_{n=1}^2 \frac{\Delta m^2_{\tilde{s}_1 
\tilde{b}_n}}{m^2_{\tilde{b}_n} - m^2_{\tilde{s}_1}}
\sum_{j=1}^4 \Biggl\{ \cr
&& \left|s_W Q_d N'_{j1} - 
\frac{1}{c_W}(1/2+ Q_d s_W^2) N'_{j2} \right|^2 
B^*_{n1} \left[ \frac{F_2(x_{jn}^b)}{m^2_{\tilde{b}_n}} - 
\frac{F_2(x_{j1}^s)}{m^2_{\tilde{s}_1}} \right] \nonumber \\
& - & \left( s_W Q_d N'_{j1} - \frac{1}{c_W} (1/2 + Q_d s_W^2) 
N'_{j2} \right) \left[ \left( s_W Q_d N'_{j1}-\frac{s_W^2}{c_W} Q_d 
N'_{j2} \right) B^*_{n2} \right. \nonumber \\
&& \qquad\qquad\left.\left.
   - \frac{m_b}{2M_W\cos\beta} N_{j3} B^*_{n1} \right] 
\frac{M_{\chi_j^0}}{m_b} 
\left[ \frac{F_4(x_{jn}^b)}{m^2_{\tilde{b}_n}} -
\frac{F_4(x_{j1}^s)}{m^2_{\tilde{s}_1}} \right] \right\}  
\label{admlr}\\
\tilde{A}_{LR}^{\gamma\Delta m_{\tilde{g}}} & = & 
- 2 \alpha_s Q_d C(R) \sum_{n=1}^2 
\frac{\Delta m^2_{\tilde{s}_1 \tilde{b}_n}}{m^2_{\tilde{b}_n} -
m^2_{\tilde{s}_1}} \nonumber \\
& \times &  \left[ B_{n1}^* \left( 
\frac{F_2(x_{\tilde gn}^b)}{m^2_{\tilde{b}_n}} -
\frac{F_2(x_{\tilde g1}^s)}{m^2_{\tilde{s}_1}} \right) - B_{n2}^* 
\frac{M_{\tilde{g}}}{m_b} \left(
\frac{F_4(x_{\tilde gn}^b)}{m^2_{\tilde{b}_n}} -
\frac{F_4(x_{\tilde g1}^s)}{m^2_{\tilde{s}_1}} \right) \right]
 \;\; \\
\tilde{A}_{LR}^{g\Delta m_{\tilde{g}}} & = &
- \alpha_s \sum_{n=1}^2 
\frac{\Delta m^2_{\tilde{s}_1 \tilde{b}_n}}{m^2_{\tilde{b}_n} -
m^2_{\tilde{s}_1}} \left\{ -B_{n1}^*C(G) \left(
\frac{F_1(x_{\tilde gn}^b)}{m^2_{\tilde{b}_n}} -
\frac{F_1(x_{\tilde g1}^s)}{m^2_{\tilde{s}_1}} \right) \right. 
\nonumber \\
&& +B_{n1}^*(2C(R)-C(G))\left(
\frac{F_2(x_{\tilde gn}^b)}{m^2_{\tilde{b}_n}} -
\frac{F_2(x_{\tilde g1}^s)}{m^2_{\tilde{s}_1}} \right) \cr
&& +B_{n2}^* \frac{M_{\tilde{g}}}{m_b} C(G)
\left( \frac{F_3(x_{\tilde gn}^b)}{m^2_{\tilde{b}_n}} -
\frac{F_3(x_{\tilde g1}^s)}{m^2_{\tilde{s}_1}} \right) \cr
&&
\left.
- B_{n2}^* \frac{M_{\tilde{g}}}{m_b} (2C(R)-C(G)) \left(
\frac{F_4(x_{\tilde gn}^b)}{m^2_{\tilde{b}_n}} -
\frac{F_4(x_{\tilde g1}^s)}{m^2_{\tilde{s}_1}} \right) \right\}
\;\; ,
\er
where $\alpha_s$ is the strong gauge coupling constant. Also,
\be
x^c_{j1}=\frac{M^2_{\chi_j^-}}{m_{\tilde{c}_1}^2}, \quad
x^s_{j1}=\frac{M^2_{\chi_j^0}}{m_{\tilde{s}_1}^2}, \quad
x^s_{\tilde g1}=\frac{M^2_{\tilde g}}{m_{\tilde{s}_1}^2}
\ee
\be
x^t_{jn}=\frac{M^2_{\chi_j^-}}{m_{\tilde{t}_n}^2}, n=1,2 \quad 
x^b_{jn}=\frac{M^2_{\chi_j^0}}{m_{\tilde{b}_n}^2}, n=1,2, \quad
x^b_{\tilde gn}=\frac{M^2_{\tilde g}}{m_{\tilde{b}_n}^2}, 
\ee
and the QCD factors are $C(R)=4/3$, $C(G)=3$.

Now we turn to the RPV contributions to $\tilde{A}_{RL}$. These are 
given by:
\be
\tilde{A}^{RPV}_{RL} = \tilde{A}_{RL}^{\lp{}} + \tilde{A}_{RL}^{\lpp{}}
+ \tilde{A}_{RL}^{\Delta m_{\chi^0}} + 
\tilde{A}_{RL}^{\Delta m_{\tilde{g}}} \; ,
\ee
with:
\br
\tilde{A}_{RL}^{\gamma,g\lp{}} & = & 
- \sum_{i,j=1}^3
\left[ \frac{\tlp{ij2}\tlp{ij3}}{4\pi} \{ Q_d,1 \} \left(
\frac{1}{12} \frac{\cos^2 \theta_{d_j}}{m^2_{\tilde{d}_j^{(1)}}}
+ \frac{1}{12} \frac{\sin^2 \theta_{d_j}}{m^2_{\tilde{d}_j^{(2)}}}
- \frac{1}{m^2_{\tilde{\nu_i}}} F_1(x_{ji}) \right) \right. 
\nonumber \\
&& +\frac{\lp{ij2}\lp{ij3}}{4\pi} \left(
-\frac{\cos^2 \theta_{e_i}}{m^2_{\tilde{e}_i^{(1)}}}
f_{\gamma,g}^{(1)}
\biggl(\frac{m^2_{u_j}}{m^2_{\tilde{e}_i^{(1)}}}\biggr) - 
\frac{\sin^2 \theta_{e_i}}{m^2_{\tilde{e}_i^{(2)}}} 
f_{\gamma,g}^{(1)}
\biggl(\frac{m^2_{u_j}}{m^2_{\tilde{e}_i^{(2)}}}\biggr) \right.
\nonumber \\ 
&& \left. \left. \qquad\qquad + 
\frac{\cos^2 \theta_{u_j}}{m^2_{\tilde{u}_j^{(1)}}} f_{\gamma,g}^{(3)}
\biggl(\frac{m^2_{e_i}}{m^2_{\tilde{u}_j^{(1)}}}\biggr)+
\frac{\sin^2 \theta_{u_j}}{m^2_{\tilde{u}_j^{(2)}}} f_{\gamma,g}^{(3)}
\biggl(\frac{m^2_{e_i}}{m^2_{\tilde{u}_j^{(2)}}}\biggr)
\right) \right]
\label{alprl}
\er
where both $\tlp{}$ and $x_{ji}$ are defined after eq.~(\ref{alplr}).
\br
\tilde{A}_{RL}^{\gamma,g\lpp{}} & = & 
-2 \sum_{i,j=1}^3 \frac{\lpp{ij2}\lpp{ij3}}{4\pi} \left[ 
\frac{\sin^2 \theta_{d_j}}{m^2_{\tilde{d}_j^{(1)}}} 
f_{\gamma,g}^{(5)} 
\biggl(\frac{ m^2_{u_i} }{ m^2_{\tilde d_j^{(1)}} } \biggr )
+\frac{\cos^2 \theta_{d_j}}{m^2_{\tilde{d}_j^{(2)}}}
f_{\gamma,g}^{(5)}
\biggl(\frac{ m^2_{u_i} }{ m^2_{\tilde d_j^{(2)}} } \biggr )
\right. \nonumber \\
&& \qquad\qquad\left. + 
\frac{\sin^2 \theta_{u_i}}{m^2_{\tilde{u}_i^{(1)}}} f_{\gamma,g}^{(6)}
\biggl(\frac{ m^2_{d_j} }{ m^2_{\tilde u_i^{(1)}} } \biggr )
(x_{j1}^i) + \frac{\cos^2 \theta_{u_i}}{m^2_{\tilde{u}_i^{(2)}}} 
f_{\gamma,g}^{(6)} 
\biggl(\frac{ m^2_{d_j} }{ m^2_{\tilde u_i^{(2)}} } \biggr )
\right]
\label{alpprl} \\
\tilde{A}_{RL}^{\gamma\Delta m_{\chi^0}} & = &
Q_d \tilde{A}_{RL}^{g\Delta m_{\chi^0}} =  
-2 \alpha_W Q_d
\sum_{n=1}^2 \frac{\Delta m^2_{\tilde{s}_2
\tilde{b}_n}}{m^2_{\tilde{b}_n} - m^2_{\tilde{s}_2}}
\sum_{j=1}^4 \left\{ \left|s_W Q_d N'_{j1} -
\frac{s_W^2}{c_W} Q_d N'_{j2} \right|^2
 \right. \nonumber \\
& \times & B^*_{n2} \left[ \frac{F_2(x_{jn}^b)}{m^2_{\tilde{b}_n}} -
\frac{F_2(x_{j2}^s)}{m^2_{\tilde{s}_2}} \right] - \left(
s_W Q_d N'^{*}_{j1} - \frac{s_W^2}{c_W} Q_d N'^{*}_{j2} \right)
\nonumber \\
& \times &
\left[ \left( s_W Q_d N'^{*}_{j1}-\frac{1}{c_W}(1/2+ Q_d s_W^2) 
N'^{*}_{j2} \right) B^*_{n1} + \frac{m_b}{2M_W\cos\beta} N^*_{j3} 
B^*_{n2} \right]
\nonumber \\
& \times & \left.
\frac{M_{\chi_j^0}}{m_b}
\left[ \frac{F_4(x_{jn}^b)}{m^2_{\tilde{b}_n}} -
\frac{F_4(x_{j2}^s)}{m^2_{\tilde{s}_2}} \right] \right\} \\
\tilde{A}_{RL}^{\gamma\Delta m_{\tilde{g}}} & = &
-2\alpha_s Q_d C(R) \sum_{n=1}^2
\frac{\Delta m^2_{\tilde{s}_2 \tilde{b}_n}}{m^2_{\tilde{b}_n} -
m^2_{\tilde{s}_2}} \nonumber \\
& \times &  \left[ B^*_{n2} \left(
\frac{F_2(x_{\tilde gn}^b)}{m^2_{\tilde{b}_n}} -
\frac{F_2(x_{\tilde g2}^s)}{m^2_{\tilde{s}_2}} \right) - B_{n1}^*
\frac{M_{\tilde{g}}}{m_b} \left(
\frac{F_4(x_{\tilde gn}^b)}{m^2_{\tilde{b}_n}} -
\frac{F_4(x_{\tilde g2}^s)}{m^2_{\tilde{s}_2}} \right) \right] \\
\tilde{A}_{RL}^{g\Delta m_{\tilde{g}}} & = &
-\alpha_s \sum_{n=1}^2
\frac{\Delta m^2_{\tilde{s}_2 \tilde{b}_n}}{m^2_{\tilde{b}_n} -
m^2_{\tilde{s}_2}} \left\{ -B^*_{n2}C(G) \left(
\frac{F_1(x_{\tilde gn}^b)}{m^2_{\tilde{b}_n}} -
\frac{F_1(x_{\tilde g2}^s)}{m^2_{\tilde{s}_2}} \right) \right.
\nonumber \\ 
&& \qquad\qquad + B_{n2}^*(2C(R)-C(G)) \left(
\frac{F_2(x_{\tilde gn}^b)}{m^2_{\tilde{b}_n}} -
\frac{F_2(x_{\tilde g2}^s)}{m^2_{\tilde{s}_2}} \right)
\nonumber \\
&& \qquad\qquad + B_{n1}^* \frac{M_{\tilde{g}}}{m_b} C(G) \left(
\frac{F_3(x_{\tilde gn}^b)}{m^2_{\tilde{b}_n}} -
\frac{F_3(x_{\tilde g2}^s)}{m^2_{\tilde{s}_2}} \right) \cr
&& \qquad\qquad \left. - B_{n1}^* \frac{M_{\tilde{g}}}{m_b} (2C(R)-C(G))
\left(
\frac{F_4(x_{\tilde gn}^b)}{m^2_{\tilde{b}_n}} -
\frac{F_4(x_{\tilde g2}^s)}{m^2_{\tilde{s}_2}} \right)
\right\}
\er

\subsection{Analytical Discussion}
In general the results for the implications of the MSSM for $b\to
s\gamma$ are well known \cite{bsgsusy}. The SM contribution
to the branching rate is large and in good agreement with the data at
around $3\times10^{-4}$. The MSSM charged Higgs contribution is always
of the same sign as that from the $W$ and can easily push the total
over the experimental limit, hence giving quite a tight constraint on
the charged Higgs mass or equivalently $m_A$. In
practice, however, the situation changes when the full spectrum is
considered, because the SUSY partners can give contributions to the
amplitude of both signs with the dominant contribution being that of
the chargino which can be of opposite sign to that of the SM and
charged Higgs \cite{bsgsusy}. There are thus two separate scenarios
according to the sign of $\mu$. For $\mu_4<0$, the structure of the
chargino mass matrix and mixings makes its contribution comparable in
size to the other two with the opposite sign giving rise to a strong
cancellation, and hence the constraints on SUSY parameters are rather
weak. For $\mu_4>0$ the chargino amplitude is not so big and moreover
its sign is not always opposed to that of SM and charged Higgs, and
therefore cannot cancel them off completely so we must have a
fairly heavy charged Higgs, which given our unification and radiative
electroweak symmetry breaking assumptions implies fairly large soft
masses.

If we now turn to the implications of $R$-parity violating couplings,
there are three distinct possibilities. The first of these is where the
product $\lp{i2j}\lp{i3j}$ is non-zero. Here the main effects will be
to change $\tilde A_{LR}$, since mixing is generated in left handed
superfields only. This will give an additional term which is added
directly to the amplitude from the usual MSSM terms, but may be of
opposite sign and hence less restricted. By contrast, the cases of
non-zero $\lp{ij2}\lp{ij3}$ or $\lpp{ij2}\lpp{ij3}$ generate
contributions to $\tilde A_{RL}$ (negligible in the MSSM), which cannot
therefore have interference with the MSSM effects but which are only
added to them in quadrature.

Beginning with the direct $R$--Parity
violating contributions, the SM contribution to $\tilde A_{LR}$ is
$-6.7\times 10^{-8}\hbox{GeV}^{-2}$, with an experimental limit on 
$\sqrt{(\tilde A_{LR}^{\gamma})^2+(\tilde A_{RL}^{\gamma})^2}$ derived
from equation~(\ref{bsglimit}) of 4.1 to 8.5$\times
10^{-8}\hbox{GeV}^{-2}$. Since the QCD corrections typically give a
contribution to $\tilde A_{i}$ of around $0.65\tilde A_i^{\gamma}$, we
shall consider what values of the couplings give a contribution of
$-2\times 10^{-8}\hbox{GeV}^{-2}$ to $\tilde A_{LR}^{\gamma}$ or
$7\times 10^{-8}\hbox{GeV}^{-2}$ to $\tilde A_{RL}^{\gamma}$ (recall
that the sign of the former matters, since it is being added to a
negative SM contribution, while the latter is added in quadrature). It
is then straightforward to derive the bounds
\br
\tlp{i2j}\tlp{i3j}(M_Z)&\lesssim& 0.09 \Biggl [
  2 \left ( \frac{100\hbox{GeV}}{m_{\tilde \nu^i}} \right )^2
  - \left ( \frac{100\hbox{GeV}}{m_{\tilde d_R^j}} \right )^2
 \Biggr ]^{-1}
\\
\vert \lp{ij2}\lp{ij3}(M_Z)\vert &\lesssim& 0.035 \Biggl [
    \left ( \frac{100\hbox{GeV}}{m_{\tilde e_L^i}} \right )^2
  - \left ( \frac{100\hbox{GeV}}{m_{\tilde d_L^j}} \right )^2
 \Biggr ]^{-1}
\\
\vert \lpp{i2j}\lpp{i3j}(M_Z)\vert &\lesssim& 0.16
    \left ( \frac{m_{\tilde q_R}}{100\hbox{GeV}} \right )^2
\label{bsglppRRdirect}
\er
where we have neglected the distinction between $\tlp{}$ and $\lp{}$ in
deriving the second of these.

The indirect contributions are far more complicated, but it is
instructive to consider their size in the case where we include only
the diagrams with a helicity flip on the gaugino line (and hence an
overall factor of the gaugino mass), set all sparticle masses
approximately degenerate at $\tilde m$ and ignore the various mixing
factors. This then gives an approximate contribution for each chargino
of
\be
\tilde{A}_{LR}^{\gamma\Delta m_{\chi^-}} \sim
 \lp{i2j}\lp{i3j}(M_{\rm GUT}) \frac{10^{-7}\hbox{GeV}^{-2}}{\cos\beta}
    \left ( \frac{100\hbox{GeV}}{\tilde m} \right )^2 
    \frac{M_{\chi_j^-}}{M_W}
\ee
where we have used equation~(\ref{dmapp}) and have evaluated the
functions with all masses equal. Since the product of two of these
couplings at the GUT scale is an order of magnitude smaller than at
$M_Z$ \cite{paper1}, for $\lp{i2j}\lp{i3j}$ we expect a significant
chargino contribution relative to the direct contributions for large
$\tan\beta$ unless the gaugino-higgsino mixing is small. A simple way
of understanding how the chargino contribution must be relevant is to
consider this in terms of results from the MSSM. The chargino
contribution here is driven by the CKM matrix element $K_{23}\simeq
0.04$, while the $R$--Parity Violating contributions are driven by a
similar mixing $\Delta m^2/m_{\tilde q}^2$, which is roughly equal to the
product of $\lp{i2j}\lp{i3j}(M_{GUT})$ and hence will be similarly
important when this product of couplings at the GUT scale is around
$10^{-2}$.

The gluino contribution is simpler, and neglecting mixing between
sbottom squarks and the mass difference between squarks and gluinos we
find a bound on each of the pairs of couplings
\br
\lp{ij2}\lp{ij3}(M_Z)&\lesssim& 0.003
    \left ( \frac{\tilde m}{100\hbox{GeV}} \right )^2
\\
\lp{ij2}\lp{ij3}(M_Z)&\lesssim& 0.006
    \left ( \frac{\tilde m}{100\hbox{GeV}} \right )^2
\\
\lpp{i2j}\lpp{i3j}(M_Z)&\lesssim& 0.006
    \left ( \frac{\tilde m}{100\hbox{GeV}} \right )^2
\er
where the bound is in fact on the sum of these indirect terms plus the
direct contributions quoted above. Unfortunately, we might expect the
suppression from the mixing in the sbottom sector to weaken these
by an order of magnitude or so. We can conclude however
that it is likely that the $RL$ contributions are dominated by the
gluino, especially in the case of non-zero $\lpp{ij2}\lpp{ij3}$, while
for $\lp{i2j}\lp{i3j}$ the chargino is likely to dominate, results
which are confirmed by our numerical studies.

\subsection{Numerical Results}
Given the discussion above, we are now in a position to
discuss the effects of setting these products non-zero with a realistic
spectrum from universality at the GUT scale. The products of couplings
which we are considering have not been constrained before, but we find
$\lp{i2j}\lp{i3j}<0.06(100\hbox{GeV}/\tilde m)^2$ using the bounds on
each coupling independently from $K^+$ decays \cite{ag}. Other such
bounds can always be evaded by selecting the indices in the products
appropriately; for example the tight bounds on $\lpp{ijk}$ from neutron
anti-neutron oscillation \cite{barbm,gs} only apply if $i=1$.

In Figure~\ref{bsglpLL} we show a simple example of mixing involving
the product $\lp{i2j}\lp{i3j}$, where we choose both couplings 0.05 at
the GUT scale, hence making the product around $0.023$ at low energy,
and with other parameters $\tan\beta=10$, $\mu_4<0$, $m_0=100$ GeV,
$A_0=0$, and $M_{1/2}$ varying. We see that in fact here the gluino
contribution is dominated by the direct one, and this in turn is
dominated by the chargino contribution. However, even at very low
values of the mass spectrum (for $M_{1/2}\lesssim 110$ GeV the chargino
mass is actually below the experimental limit) and even though we have
taken the product of couplings to be close to the limit of
equation~(\ref{sneutmassbound}), the contributions are still too small to
greatly affect the total. Hence we conclude that bounds on this product
are really not usefully found from $b\to s\gamma$ decays.

Similarly, for $\lp{ij2}\lp{ij3}$ we expect that the product at the GUT
scale will be $\lesssim 0.01$ from equation~(\ref{sneutmassbound}), which
in turn leads to the product at the weak scale being around $\lesssim
0.1$, which can only give a useful bound if the spectrum is extremely
light, and in practice it again appears that constructing a spectrum so
light as to have interesting consequences for $b\to s\gamma$ through
$R$-Parity Violation requires one or other of the masses to become
lighter than its experimental bound.

For the case of $\lpp{ij2}\lpp{ij3}$ we do not have such tight bounds
from the sneutrino mass and can increase the coupling to quite a large
value. In Figure~\ref{bsglppRR}, we show a plot with the same input
parameters as in Figure~\ref{bsglpLL}, but with
$\lpp{ij2}\lpp{ij3}(M_{GUT})=0.01$. Here again the effects on $b\to
s\gamma$ are only really beginning to be significant for a very light
spectrum. However, it is interesting that the contribution is very much
larger than might be expected from equation~(\ref{bsglppRRdirect}), since
the gluino term is overwhelmingly dominant.

In conclusion, while the $R$--Parity violating couplings give a
contribution to $b\to s\gamma$ which is enhanced typically by an order
of magnitude when the indirect contributions are included, and although
this in principle gives a tighter bound than others in the literature,
in fact these constraints are very weak in the context of universal
mass spectrum at the GUT scale. This is because, for example, it is
virtually impossible to arrange that squarks should have masses as
light as 100 GeV without violating one or another of the experimental
limits. In the case of non-zero $\lp{}$ couplings the bounds from $b\to
s\gamma$ are weaker than those from the sneutrino mass limits given in
equation~(\ref{sneutmassbound}). For the $\lpp{}$ couplings, the
indirect gluino contribution is far larger than the direct
contribution, and we find that for a light spectrum with $m_0$ and
$M_{1/2}$ of order 100 GeV, the product of $\lpp{ij2}\lpp{ij3}(M_Z)$
should be $\lesssim 0.2$. This bound scales very roughly as $M_{1/2}$,
but with some complicated dependence on the sbottom squark mixing as
well as on the mass spectrum.

\section{$K^0-\bar K^0$ mixing}
\subsection{SUSY contributions}
The $K^0-\bar K^0$ sector has long been a probe of physics beyond the
standard model, starting with the original motivation of the study of
CP violation and recently also as a way of investigating flavour
changing neutral currents in SUSY. These occur even in the absence of
$R$-Parity violation, since in the MSSM (and SM) quark flavour is not
conserved. Hence we shall find that the RPV contribution is
complementary to the already existing CKM contribution. We shall ask
the question of how much it is possible to constrain RPV couplings
through their impact on $\Delta m_K$.  Here we shall consider the
relative sizes of three effects, the MSSM $\Delta m^2$ contribution and
the direct and indirect RPV contributions. The first has already been
investigated \cite{hagelin,otherfcnc}, as has the last
\cite{barbm,bgnn,ag,crs,chr}, but the question of whether the simple
direct effects are in fact dominant has never been studied.

The upper bound on the $K_L-K_S$ mass difference has been
measured \cite{pdb} as
\be
\Delta m_K = (3.491 \pm 0.009) \times 10^{-15} \hbox{GeV}
\ee
$K^0-\bar K^0$ mixing is generated by the effective Lagrangian
\br
\Delta{\cal L}^{\Delta S=2}&=&
c_{LL}[\bar d_i P_L s^i][\bar d_j P_L s^j]
+c^{\prime}_{LL}[\bar d_i P_L s^j][\bar d_j P_L s^i]\cr
&+&
c_{RR}[\bar d_i P_R s^i][\bar d_j P_R s^j]
+c^{\prime}_{RR}[\bar d_i P_R s^j][\bar d_j P_R s^i]\cr
&+&
c_{LR}[\bar d_i P_L s^i][\bar d_j P_R s^j]
+c^{\prime}_{LR}[\bar d_i P_L s^j][\bar d_j P_R s^i]\cr
&+&
d_{LL}[\bar d_i \gamma_\mu P_L s^i][\bar d_j \gamma^\mu P_L s^j]
+d_{RR}[\bar d_i \gamma_\mu P_R s^i][\bar d_j \gamma^\mu P_R s^j]
\er
giving a contribution to the $K_L-K_S$ mass difference
\br
\Delta m_K&=&\frac{1}{12} f_K^2m_K
  \Bigl (
   c_{LR} + 3c^{\prime}_{LR} + 8d_{LL} +8d_{RR} \cr
&& \qquad
 + \frac{m_K^2}{(m_d+m_s)^2}
 (6c_{LR}+2c^{\prime}_{LR}+5c_{LL}-c^{\prime}_{LL}
            +5c_{RR}-c^{\prime}_{RR})
  \Bigr )
\label{dmKeqn}
\er
Here the terms involving both left and right handed fields in the
effective lagrangian will give a larger contribution to $\Delta m_K$
than those involving purely left or right handed fields because of the
factor of $m_K^2/m_s^2$. We shall not try to include QCD corrections to
these formulae since these are unlikely to be more significant than the
errors implicit in, for example, our choice of $\alpha_3$, $m_s$, and
SUSY parameters.

Contributions to the parameters in the effective Lagrangian come from a
variety of different classes of diagram, some of which are shown in
Figure~\ref{KKdiag}. The standard model contribution from
Figure~\ref{KKdiag}a is
\br
{d_{LL}}^{SM}&=&\sum_{i,j}\frac{g^4}{64\pi^2}
   K_{i1}K^*_{i2}K_{j2}K^*_{j1} I_4(m_{u_i}^2,m_{u_j}^2,M_W^2,M_W^2) \cr
&\simeq& \frac{g^4}{128\pi^2}\vert K_{cd}\vert^2\frac{m_c^2}{M_W^4}
\label{SMKKbar}
\er
where for the second line we only consider the charm and up quark
contributions.

Apart from the standard model contributions, there are contributions
from direct $R$-Parity violating diagrams which have been calculated
previously \cite{barbm,ag,crs,chr}. The only tree level diagram is that
with the interchange of a sneutrino and two $\tlp{}$ couplings shown in
Figure~\ref{KKdiag}b
\be
{c_{LR}}^{TL}=-\sum_i\frac{ \tlp{i21}\tlp{i12}^* }
                      { m_{\tilde\nu_i}^2 }
\ee
where our notation is as in the previous section and our earlier work
\cite{paper1}.

The $\lp{}$ box diagrams such as Figure~\ref{KKdiag}c give a
contribution
\br
{d_{LL}}^{\lp{}}&=&\sum_{i,j,k,m}
 \frac{1}{64\pi^2}\tlp{i1k}^*\tlp{j2k}\tlp{j1m}^*\tlp{i2m}
      I_4(m_{\tilde\nu_i}^2,m_{\tilde\nu_j}^2,m_{d_k}^2,m_{d_m}^2) \cr
&+&\sum_{i,j,k,m}
 \frac{1}{64\pi^2}\tlp{i1k}^*\tlp{j2k}\tlp{j1m}^*\tlp{i2m}
      I_4(m_{\nu_i}^2,m_{\nu_j}^2,m_{\tilde d_R^k}^2,m_{\tilde d_R^m}^2)\\
{d_{RR}}^{\lp{}}&=&\sum_{i,j,k,m}
 \frac{1}{64\pi^2}\tlp{ik1}\tlp{jk2}^*\tlp{jm1}\tlp{im2}^*
      I_4(m_{\tilde\nu_i}^2,m_{\tilde\nu_j}^2,m_{d_k}^2,m_{d_m}^2) \cr
&+&\sum_{i,j,k,m}
 \frac{1}{64\pi^2}\tlp{ik1}\tlp{jk2}^*\tlp{jm1}\tlp{im2}^*
      I_4(m_{\nu_i}^2,m_{\nu_j}^2,m_{\tilde d_L^k}^2,m_{\tilde d_L^m}^2)\cr
&+&\sum_{i,j,k,m}
 \frac{1}{64\pi^2}\lp{ik1}\lp{jk2}^*\lp{jm1}\lp{im2}^*
      I_4(m_{\tilde e_L^i}^2,m_{\tilde e_L^j}^2,m_{u_k}^2,m_{u_m}^2) \cr
&+&\sum_{i,j,k,m}
 \frac{1}{64\pi^2}\lp{ik1}\lp{jk2}^*\lp{jm1}\lp{im2}^*
      I_4(m_{e_i}^2,m_{e_j}^2,m_{\tilde u_L^k}^2,m_{\tilde u_L^m}^2)\\
{c^{\prime}_{LR}}^{\lp{}}&=&-\sum_{i,j,k,m}
 \frac{1}{32\pi^2}\tlp{i1k}^*\tlp{j2k}\tlp{im1}\tlp{jm2}^*
      I_4(m_{\tilde\nu_i}^2,m_{\tilde\nu_j}^2,m_{d_k}^2,m_{d_m}^2) \cr
&&-\sum_{i,j,k,m}
 \frac{1}{32\pi^2}\tlp{i1k}^*\tlp{j2k}\tlp{im1}\tlp{jm2}^*
      I_4(m_{\nu_i}^2,m_{\nu_j}^2,m_{\tilde d_R^k}^2,m_{\tilde d_R^m}^2)
\er
while $\lpp{}$ box diagrams such as Figure~\ref{KKdiag}d give
\br
{d_{RR}}^{\lpp{}}&=&\sum_{i,j}
 \frac{1}{32\pi^2}\lpp{i13}\lpp{i23}^*\lpp{j13}\lpp{j23}^*
\biggl [
      I_4(m_b^2,m_b^2,m_{\tilde u_R^i}^2,m_{\tilde u_R^j}^2) \cr
&&  \qquad\qquad\qquad\qquad\qquad\qquad
    + I_4(m_{\tilde b_R}^2,m_{\tilde b_R}^2,m_{u_i}^2,m_{u_j}^2)
\biggr ]
\er

In addition we have diagrams such as those Figure~\ref{KKdiag}e with
one internal $W$ boson and one internal squark line, where a helicity
flip is needed on the internal fermion line, forcing it to be a top or
charm. Hence we have
\br
{c_{LR}}^{\lpp{}W}&=&-\sum_{i,j}
 \frac{\alpha}{4\pi\sin^2\theta_W}\lpp{i31}\lpp{j32}^*K_{i1}^*K_{j2}
      m_{u_i}m_{u_j}J_4(m_{\tilde b_R}^2,M_W^2,m_{u_i}^2,m_{u_j}^2) \cr
{c^{\prime}_{LR}}^{\lpp{}W}&=&\sum_{i,j}
 \frac{\alpha}{4\pi\sin^2\theta_W}\lpp{i31}\lpp{j32}^*K_{i1}^*K_{j2}
      m_{u_i}m_{u_j}J_4(m_{\tilde b_R}^2,M_W^2,m_{u_i}^2,m_{u_j}^2)
\er
We neglect similar diagrams with $\lp{}$ vertices, since
these are always dominated by the tree level contribution.

Although they have not been considered in the context of $R$-Parity,
the diagrams involving mass insertions on squark lines have been
analysed in the case where the off-diagonal mass terms are generated by
the CKM matrix and by boundary conditions at the unification scale
\cite{hagelin,otherfcnc,bert}. Most of these analyses include only
gluino mediated box diagrams, such as Figure~\ref{KKdiag}f.
\br
{d_{LL}}^{\tilde g}&=&
\left( \Delta m^2_{\tilde s_L\tilde d_L}\right)^2
  \alpha_3^2 \biggl [
\frac{11}{18} I_4^{\prime\prime}
 (m_{\tilde d_L}^2,m_{\tilde d_L}^2,M_{\tilde g}^2,M_{\tilde g}^2) \cr
&&\qquad\qquad\qquad\qquad
-\frac{1}{9} M_{\tilde g}^2 J_4^{\prime\prime}
 (m_{\tilde d_L}^2,m_{\tilde d_L}^2,M_{\tilde g}^2,M_{\tilde g}^2)
\biggr ] \\
{d_{RR}}^{\tilde g}&=&
\left( \Delta m^2_{\tilde s_R\tilde d_R}\right)^2
  \alpha_3^2 \biggl [
\frac{11}{18} I_4^{\prime\prime}
 (m_{\tilde d_R}^2,m_{\tilde d_R}^2,M_{\tilde g}^2,M_{\tilde g}^2) \cr
&&\qquad\qquad\qquad\qquad
-\frac{1}{9} M_{\tilde g}^2 J_4^{\prime\prime}
 (m_{\tilde d_R}^2,m_{\tilde d_R}^2,M_{\tilde g}^2,M_{\tilde g}^2)
\biggr ] \\
{c_{LR}}^{\tilde g}&=&
\left(\Delta m^2_{\tilde s_L\tilde d_L}\right)
\left(\Delta m^2_{\tilde s_R\tilde d_R}\right)
  \alpha_3^2 \biggl [
-\frac{2}{3} I_4^{\prime\prime}
 (m_{\tilde d_L}^2,m_{\tilde d_R}^2,M_{\tilde g}^2,M_{\tilde g}^2) \cr
&&\qquad\qquad\qquad\qquad\qquad
-\frac{7}{3} M_{\tilde g}^2 J_4^{\prime\prime}
 (m_{\tilde d_L}^2,m_{\tilde d_R}^2,M_{\tilde g}^2,M_{\tilde g}^2)
\biggr ] \\
&+&
\left(\Delta m^2_{\tilde s_L\tilde d_R}\right)
\left(\Delta m^2_{\tilde s_R\tilde d_L}\right)
  \alpha_3^2 \biggl [
-\frac{11}{9} I_4^{\prime\prime}
 (m_{\tilde d_{L}}^2,m_{\tilde d_{R}}^2,M_{\tilde g}^2,M_{\tilde g}^2) 
 \biggr ] \cr
{c_{LR}^{\prime}}^{\tilde g}&=&
\left(\Delta m^2_{\tilde s_L\tilde d_L}\right)
\left(\Delta m^2_{\tilde s_R\tilde d_R}\right)
  \alpha_3^2 \biggl [
\frac{10}{9} I_4^{\prime\prime}
(m_{\tilde d_L}^2,m_{\tilde d_R}^2,M_{\tilde g}^2,M_{\tilde g}^2) \cr
&& \qquad\qquad\qquad\qquad\qquad
-\frac{1}{9} M_{\tilde g}^2 J_4^{\prime\prime}
 (m_{\tilde d_L/}^2,m_{\tilde d_R}^2,M_{\tilde g}^2,M_{\tilde g}^2)
\biggr ] \\
&+&
\left(\Delta m^2_{\tilde s_L\tilde d_R}\right)
\left(\Delta m^2_{\tilde s_R\tilde d_L}\right)
  \alpha_3^2 \biggl [
-\frac{5}{3} I_4^{\prime\prime}
 (m_{\tilde d_{L}}^2,m_{\tilde d_{R}}^2,M_{\tilde g}^2,M_{\tilde g}^2)
 \biggr ] \cr
{c_{LL}}^{\tilde g}&=&
\left(\Delta m^2_{\tilde s_L\tilde d_R}\right)^2
  \alpha_3^2 \biggl [
-\frac{17}{18} M_{\tilde g}^2 J_4^{\prime\prime}
 (m_{\tilde d_{L}}^2,m_{\tilde d_{R}}^2,M_{\tilde g}^2,M_{\tilde g}^2)
\biggr ] \\
{c_{LL}^\prime}^{\tilde g}&=&
\left(\Delta m^2_{\tilde s_L\tilde d_R}\right)^2
  \alpha_3^2 \biggl [
 \frac{1}{6} M_{\tilde g}^2 J_4^{\prime\prime}
 (m_{\tilde d_{L}}^2,m_{\tilde d_{R}}^2,M_{\tilde g}^2,M_{\tilde g}^2)
\biggr ] \\
{c_{RR}}^{\tilde g}&=&
\left(\Delta m^2_{\tilde s_R\tilde d_L}\right)^2
  \alpha_3^2 \biggl [
-\frac{17}{18} M_{\tilde g}^2 J_4^{\prime\prime}
 (m_{\tilde d_{L}}^2,m_{\tilde d_{R}}^2,M_{\tilde g}^2,M_{\tilde g}^2)
\biggr ] \\
{c_{RR}^\prime}^{\tilde g}&=&
\left(\Delta m^2_{\tilde s_R\tilde d_L}\right)^2
  \alpha_3^2 \biggl [
 \frac{1}{6} M_{\tilde g}^2 J_4^{\prime\prime}
 (m_{\tilde d_{L}}^2,m_{\tilde d_{R}}^2,M_{\tilde g}^2,M_{\tilde g}^2)
\biggr ] \;\;,
\er
where we have assumed that $m_{\tilde d_L}\simeq m_{\tilde s_L}$.
These results and those for the resulting contribution to $\Delta m_K$
in equation~(\ref{dmKeqn}) agree with those presented in Gabbianni 
{\it et al} \cite{otherfcnc}, where the disagreements between these 
results and others in references~\cite{hagelin,otherfcnc} are 
discussed.

The chargino mediated box diagrams give
\br
{d_{LL}}^{\tilde\chi^{\pm}}&=&
\left( \Delta m^2_{\tilde c_L\tilde u_L}\right)^2
  \frac{\alpha^2}{4\sin^4\theta_w} \sum_{i,j} 
\vert V_{i1}\vert^2 \vert V_{j1}\vert^2
I_4^{\prime\prime} (m_{\tilde u_L}^2,m_{\tilde u_L}^2,
  M_{\tilde\chi^{\pm}_i}^2,M_{\tilde\chi^{\pm}_j}^2)
\er
while from neutralino mediated box diagrams we have
\br
{d_{LL}}^{\tilde\chi^0}&=& \sum_{ij}
\frac{1}{16\pi^2}
\left( \Delta m^2_{\tilde s_L\tilde d_L}\right)^2
\biggl [
 \vert A_{dLi} \vert^2 \vert A_{dLj} \vert^2 I_4^{\prime\prime}
 (m_{\tilde d_L}^2,m_{\tilde d_L}^2,M_{\tilde\chi_i^0}^2,M_{\tilde\chi_j^0}^2) 
\cr
&&\qquad\qquad -(A_{dLi})^2 ( A_{dLj}^* )^2 
 M_{\tilde\chi_i^0}M_{\tilde\chi_j^0} J_4^{\prime\prime}
 (m_{\tilde d_L}^2,m_{\tilde d_L}^2,M_{\tilde\chi_i^0}^2,M_{\tilde\chi_j^0}^2)
\biggr ] \\
{d_{RR}}^{\tilde\chi^0}&=& \sum_{ij}
\frac{1}{16\pi^2}
\left( \Delta m^2_{\tilde s_R\tilde d_R}\right)^2
\biggl [
 \vert A_{dRi} \vert^2 \vert A_{dRj} \vert^2 I_4^{\prime\prime}
 (m_{\tilde d_R}^2,m_{\tilde d_R}^2,
                       M_{\tilde\chi_i^0}^2,M_{\tilde\chi_j^0}^2) \cr
&&\qquad\qquad -(A_{dRi})^2 ( A_{dRj}^* )^2 
 M_{\tilde\chi_i^0}M_{\tilde\chi_j^0} J_4^{\prime\prime}
 (m_{\tilde d_R}^2,m_{\tilde d_R}^2,M_{\tilde\chi_i^0}^2,M_{\tilde\chi_j^0}^2)
\biggr ]\\
{c_{LR}}^{\tilde\chi^0}&=& -\sum_{ij}
\frac{ A_{dLi}A_{dRi}^*A_{dLj}^*A_{dRj} }
     {4\pi^2}
\left(\Delta m^2_{\tilde s_L\tilde d_L}\right)
\left(\Delta m^2_{\tilde s_R\tilde d_R}\right) \cr
&& \qquad\qquad\qquad\qquad
\times
M_{\tilde\chi_i^0}M_{\tilde\chi_j^0} J_4^{\prime\prime}
 (m_{\tilde d_L}^2,m_{\tilde d_R}^2,M_{\tilde\chi_i^0}^2,M_{\tilde\chi_j^0}^2)\\
{c_{LR}^{\prime}}^{\tilde\chi^0}&=& \sum_{ij}
\frac{ A_{dLi}A_{dRi}A_{dLj}^*A_{dRj}^* }
     {4\pi^2}
\left(\Delta m^2_{\tilde s_L\tilde d_L}\right)
\left(\Delta m^2_{\tilde s_R\tilde d_R}\right) \cr
&& \qquad\qquad\qquad\qquad
\times I_4^{\prime\prime}
 (m_{\tilde d_L}^2,m_{\tilde d_R}^2,M_{\tilde\chi_i^0}^2,M_{\tilde\chi_j^0}^2)
\er
and from those with one gluino and one neutralino :
\br
{d_{LL}}^{\tilde\chi^0\tilde g}&=& -\sum_{i}
\frac{\alpha_3}{6\pi}
\left( \Delta m^2_{\tilde s_L\tilde d_L}\right)^2
\biggl [
\vert A_{dLi} \vert^2 I_4^{\prime\prime}
 (m_{\tilde d_L}^2,m_{\tilde d_L}^2,M_{\tilde\chi_i^0}^2,M_{\tilde g}^2) \cr
&&\qquad\qquad
+\frac{1}{2}(A_{dLi}^{*2} + A_{dLi}^2)
M_{\tilde\chi_i^0}M_{\tilde g} J_4^{\prime\prime}
 (m_{\tilde d_L}^2,m_{\tilde d_L}^2,M_{\tilde\chi_i^0}^2,M_{\tilde g}^2)
\biggr ] \\
{d_{RR}}^{\tilde\chi^0\tilde g}&=& -\sum_{i}
\frac{\alpha_3}{6\pi}
\left( \Delta m^2_{\tilde s_R\tilde d_R}\right)^2
\biggl [
\vert A_{dRi} \vert^2 I_4^{\prime\prime}
 (m_{\tilde d_R}^2,m_{\tilde d_R}^2,M_{\tilde\chi_i^0}^2,M_{\tilde g}^2) \cr
&&\qquad\qquad
+\frac{1}{2}(A_{dRi}^{*2} + A_{dRi}^2)
M_{\tilde\chi_i^0}M_{\tilde g} J_4^{\prime\prime}
 (m_{\tilde d_R}^2,m_{\tilde d_R}^2,M_{\tilde\chi_i^0}^2,M_{\tilde g}^2)
\biggr ] \\
\label{dRRglu}
{c_{LR}}^{\tilde\chi^0\tilde g}&=& -\sum_{i}
\frac{\alpha_3}{2\pi}
\left(\Delta m^2_{\tilde s_L\tilde d_L}\right)
\left(\Delta m^2_{\tilde s_R\tilde d_R}\right)
\biggl [
(A_{dRi}A_{dLi} + A_{dRi}^* A_{dLi}^* ) \cr
&& \qquad\qquad\qquad\qquad\qquad\qquad\qquad \times I_4^{\prime\prime}
 (m_{\tilde d_L}^2,m_{\tilde d_R}^2,M_{\tilde\chi_i^0}^2,M_{\tilde g}^2) \cr
&&\qquad\qquad
+ (A_{dRi}A_{dLi}^* + A_{dRi}^* A_{dLi} )
M_{\tilde\chi_i^0}M_{\tilde g} J_4^{\prime\prime}
 (m_{\tilde d_L}^2,m_{\tilde d_R}^2,M_{\tilde\chi_i^0}^2,M_{\tilde g}^2)
\biggr ] \\
{c_{LR}^{\prime}}^{\tilde\chi^0\tilde g}&=& \sum_{i}
\frac{\alpha_3}{6\pi}
\left(\Delta m^2_{\tilde s_L\tilde d_L}\right)
\left(\Delta m^2_{\tilde s_R\tilde d_R}\right)
\biggl [
(A_{dRi}A_{dLi} + A_{dRi}^* A_{dLi}^* ) \cr
&& \qquad\qquad\qquad\qquad\qquad\qquad\qquad \times I_4^{\prime\prime}
 (m_{\tilde d_L}^2,m_{\tilde d_R}^2,M_{\tilde\chi_i^0}^2,M_{\tilde g}^2) \cr
&&\qquad\qquad
+ (A_{dRi}A_{dLi}^* + A_{dRi}^* A_{dLi} )
M_{\tilde\chi_i^0}M_{\tilde g} J_4^{\prime\prime}
 (m_{\tilde d_L}^2,m_{\tilde d_R}^2,M_{\tilde\chi_i^0}^2,M_{\tilde g}^2)
\biggr ] \qquad
\er
Here we use
\br
A_{dLj}&=&eQ_dN_{j1}^{\prime} - \frac{g}{\cos\theta_W}
  \left( \frac{1}{2}+Q_d\sin^2\theta_W\right)N_{j2}^{\prime}\cr
A_{dRj}&=&-eQ_dN_{j1}^{\prime*}
 + \frac{gQ_d\sin^2\theta_W}{\cos\theta_W}N_{j2}^{\prime*}
\er
and we note that
\br
\Delta m^2_{\tilde s_L \tilde d_L}&=&m^2_{Q_1Q_2}
\nonumber \\
\Delta m^2_{\tilde s_R \tilde d_R}&=&m^2_{d_1d_2}
 \\
(\Delta m^2_{\tilde s_L \tilde d_R})^2&=&\nu_1^2\left( \eta^d_{21}
\right)^2
   +\nu_1^2 \eta^d_{21} \left( \eta^d_{11}
           m^2_{Q_1Q_2}\frac{\delta}{\delta m^2_{\tilde d_L}}
   +\eta^d_{22}
           m^2_{d_1d_2}\frac{\delta}{\delta m^2_{\tilde d_R}} \right)
   + \ldots
\nonumber \\
(\Delta m^2_{\tilde s_R \tilde d_L})^2&=&\nu_1^2 \left( \eta^d_{12}
\right)^2
   +\nu_1^2 \eta^d_{12} \left( \eta^d_{22}
           m^2_{Q_1Q_2}\frac{\delta}{\delta m^2_{\tilde d_L}}
   +\eta^d_{11}
           m^2_{d_1d_2}\frac{\delta}{\delta m^2_{\tilde d_R}} \right)
   + \ldots
\nonumber \\
\Delta m^2_{\tilde s_L \tilde d_R} \Delta m^2_{\tilde s_R \tilde d_L}
&=& \nu_1^2 \left( \eta^d_{21}\eta^d_{12}+ \frac{1}{2} \eta^d_{21} 
\left(\eta^d_{22}
           m^2_{Q_1Q_2}\frac{\delta}{\delta m^2_{\tilde d_L}}
   +\eta^d_{11}
           m^2_{d_1d_2}\frac{\delta}{\delta m^2_{\tilde d_R}} \right)
 \right.  \nonumber \\
&+&
\left.   \frac{1}{2} \eta^d_{12} \left( \eta^d_{11}
           m^2_{Q_1Q_2}\frac{\delta}{\delta m^2_{\tilde d_L}}
   +\eta^d_{22}
           m^2_{d_1d_2}\frac{\delta}{\delta m^2_{\tilde d_R}} \right) 
+ \ldots \right)
\nonumber 
\er
where the ellipsis stands for higher order contributions, the
derivatives are  assumed to act only of one of the arguments of the
appropriate $I_4^{\prime\prime}$, and we have only included terms
proportional to 
$\Delta m^2_{\tilde s_L \tilde d_R}$
and 
$\Delta m^2_{\tilde s_R \tilde d_L}$
in the dominant gluino contributions, not in the neutralino
contributions where such terms also appear but are smaller.

\subsection{Analytical Discussion}
We begin by noting that the standard model term from
equation~(\ref{SMKKbar}) gives a contribution to $\Delta m_K$ of around
$2\times 10^{-15}$ GeV, which given the large theoretical errors in the
input parameters to the calculation is in reasonable agreement. Hence
we shall here require that the new contributions from SUSY and RPV do
not destroy this agreement by being larger than the experimental limit
themselves. However, it is important to note that there are unknown
relative signs between the various contributions, and hence that there
can be cancellations.

The direct contributions have been discussed in
refs.~\cite{barbm,bgnn,ag,crs,chr}. The most stringent bound on couplings
is that from the tree level diagram of Figure~\ref{KKdiag}b which leads
to a constraint \cite{bgnn,chr}
\be
\tlp{i12}\tlp{i21}(M_Z)
\equiv\lp{ij2}\lp{ik1}(M_Z)K_{j1}K_{k2}
\lesssim 1.3\times 10^{-7}
 \left( \frac{\tilde m_{\tilde\nu_i}}{500\hbox{GeV}}\right )^2
\label{TLbound}
\ee
The box diagrams of Figures~\ref{KKdiag}c and \ref{KKdiag}d and the
competing diagram with an internal W line of Figure~\ref{KKdiag}e lead
to bounds on products of two couplings of order $10^{-2}$ to $10^{-3}$
\cite{barbm,crs} for a very light spectrum. Ref.~\cite{ag} was mostly
concerned with the case where, in the weak basis, only one $R$-Parity
violating coupling is non-zero. This leads to bounds of order 0.1 on
$\lp{i1k}$ and $\lp{i2k}$ for a very light spectrum, but scaling more
weakly with masses.

We now consider how large the expected contributions from the main
indirect processes are, beginning with non-zero $\lpp{}$, by deriving
approximate bounds from each of the most important direct and indirect
contributions in turn. The term which we shall bound will be
$\lpp{}^2\equiv\lpp{i31}\lpp{i32}$, and for purposes of the discussion
in this section we shall assume that all superpartners are degenerate
at $\tilde m\simeq 3M_{1/2}$.

The main indirect contributions are the gluino mediated box diagrams,
which arise because
$\Delta m^2_{\tilde s_R\tilde d_R}$
is non-zero. Using the various functions in Appendix C and
Eqs.~(\ref{dmapp}) and (\ref{dRRglu}) we can find $d_{RR}^{\tilde g}$
and hence derive an approximate bound of
\be
\lpp{}^2(M_Z)\lesssim 0.07
 \left( \frac{\tilde m}{500\hbox{GeV}} \right )
\label{dRRcont}
\ee
Although other indirect contributions exist, this is the simplest and is
sometimes the largest. However, the various contributions from
$\Delta m^2_{\tilde s_L \tilde d_R}$ and
$\Delta m^2_{\tilde s_R \tilde d_L}$
often in practice give comparably large effects, and so
equation~(\ref{dRRcont}) should be treated with caution.

The direct contributions are those from the $\lpp{}$ box diagram and
the $\lpp{}-W$ diagram \cite{barbm,crs}. The former gives a bound of
\be
\lpp{}^2(M_Z) \lesssim 0.01
 \left( \frac{\tilde m}{500\hbox{GeV}}\right )
\label{dRRdirectcontn}
\ee
which may be rather inaccurate if the first index of the non-zero
$\lpp{}$ is 3. The second gives
\br
\lpp{213}\lpp{223}(M_Z) &\lesssim& 0.05
 \left( \frac{\tilde m}{500\hbox{GeV}}\right )^2 \\
\lpp{213}\lpp{323}(M_Z) &\lesssim& 0.1
 \left( \frac{\tilde m}{500\hbox{GeV}}\right )^2 \\
\lpp{313}\lpp{223}(M_Z) &\lesssim& 0.2
 \left( \frac{\tilde m}{500\hbox{GeV}}\right )^2 \\
\lpp{313}\lpp{323}(M_Z) &\lesssim& 0.1
 \left( \frac{\tilde m}{500\hbox{GeV}}\right )^2 \label{toptop}
\er
respectively, where the scaling is very approximate for the cases
involving stop squarks. These numbers are in reasonable agreement with
those in reference~\cite{crs} given the uncertainties in (for example)
$m_s$.

We conclude that although the indirect contribution appears rather
smaller, none of the contributions to $\Delta m_K$ is obviously
negligible in the region of interest, and so we must turn to a
numerical analysis, noting that we expect $\lpp{}^2(M_{GUT})\sim
10^{-3}$ to give a result comparable with the experimental limit for
masses of order a few hundred GeV.

We now turn to the case of non-zero $\lp{}$. Here the tree level
diagram is inevitably completely dominant where it exists
\cite{bgnn,chr}, and even with only one non-zero coupling in the weak
basis there can be measurable effects \cite{ag}. Note that here it is
possible to arrange the couplings so that the indirect contributions
are zero but the direct contribution is not, for example if we arrange
for non-zero $\tlp{113}$, $\tlp{223}$, $\tlp{212}$, and $\tlp{122}$.
However, such scenarios seem contrived given the need to avoid
the extremely tight bounds from the tree level term while
simultaneously having non-negligible values for four couplings,
requiring remarkable cancellations in $\tlp{i12}\tlp{i21}$.

The only scenario which we shall consider is thus non-zero $\lp{i13}$
and $\lp{i23}$. For this case the direct contribution lead to a
constraint which is similar to that of Eq.~(\ref{dRRcont}),
\be
\lp{}^2(M_Z)\lesssim 0.07
 \left( \frac{\tilde m}{500\hbox{GeV}} \right )
\label{dLLcont}
\ee
where again this result is very approximate given the number of
different contributions which can be relevant. There is no $\lp{}-W$
diagram by construction since it only exists when the tree level
diagram exists, while the direct contribution from the box diagram then
gives
\be
\lp{}^2(M_Z) \lesssim 8\times 10^{-3}
 \left( \frac{\tilde m}{500\hbox{GeV}}\right )
\label{dLLdirectcontn}
\ee
Since these results are in general only reliable as order of magnitude
estimates, we use a numerical study to find out which contributions are
most significant.

\subsection{Numerical Results}
The results for $\Delta m_K$ are fairly straightforward. We find that
for each scenario which we have considered the indirect contributions
are in fact completely dominated by the direct ones, with the gluino
term being between one and two orders of magnitude too small to compete
with the direct box diagrams. We illustrate this in
Figure~\ref{KKbarlppRR}, which shows a typical situation. Here we have
set $\lpp{213}(M_{GUT})=\lpp{223}(M_{GUT})=0.02$, $M_{1/2}=100$ GeV,
$A_0=0$, $\tan\beta=10$, $\mu_4<0$, and show the relevant
contributions.

Apart from the fact that the direct contributions are dominant here, we
note that the contribution from diagrams with one neutralino and one
gluino line is comparable in magnitude to that from gluinos alone. This
clearly will be significant for models involving mass insertions from a
GUT theory. In general we find that for flavour violation in the left
handed sector the contributions from charginos and mixed
gluino-neutralino diagrams are up to half that of the gluino, while for
right handed flavour violation the chargino contribution is negligible
but the mixed neutralino-gluino contribution is similar.

In summary then, we find that in practice this is the only process
which has been studied where the indirect contributions are generically
overwhelmed by the direct one, and hence we can essentially simply use
the results quoted above in equations~(\ref{dRRdirectcontn}) and
(\ref{dLLdirectcontn}) to bound the couplings, thus confirming the
results of references~\cite{bgnn,ag,crs,chr}. Why this should be so is
unclear, since the formulae for the contributions to $\Delta m_K$ are
so complicated, but it appears the values of the various four point
functions are simply such as to suppress the indirect contributions,
while for the three point functions relevant for $\mu\to e\gamma$ and
$b\to s\gamma$ the indirect contributions are enhanced. We also find
that different parts in the gluino contribution are of different signs
and comparable magnitude, and so tend to partially cancel.

Finally, we note that the qualitative nature of this result is such
that the direct contribution will also dominate in $B^0-\bar B^0$ and
$D^0-\bar D^0$ mixing.

\section{Conclusion}
We now conclude with a brief summary of our results. We have extended
our analysis of the RGEs for $R$-Parity violating supersymmetry to
include the effects of the CKM matrix, and have studied two well-known
flavour changing processes, including both the direct contributions,
with $R$-Parity violating couplings at the vertices of diagrams, and
the indirect ones, where flavour violation arises through soft masses
generated by the Renormalisation Group Equations.

We first showed that constraints on $\lp{}$ can be derived by demanding
that the sneutrino masses not be driven below their experimental
limits, which gives a bound of around 0.3 on the low energy values of
the $\lp{}$ couplings, with the particularly interesting feature that
it is extremely insensitive to the sparticle spectrum and does not
disappear in the limit where the masses go to infinity. Given that the
sparticle spectrum is now known to be heavy (with many limits well
above 100 GeV if we assume unification) this means that this bound is
one of the tightest existing in such a scenario.

For the case of $b\to s\gamma$ we find that the indirect contributions
caused by the flavour violation are dominant, and enhance the amplitude
contribution by up to an order of magnitude, allowing new constraints
on both $\lpp{}$ and $\lp{}$. However, these are rather weak, and the
$\lp{}$ constraints are weaker than those from sneutrino masses derived
earlier in this paper.

With regard to $K^0-\bar K^0$ mixing, we have included all the indirect
contributions including those from gluino, neutralino, chargino, and
mixed neutralino-gluino diagrams. We find that here the chargino and
mixed diagrams are sometimes comparable in size to those from gluinos,
with consequences for the study of flavour changing from GUT theories.
However, in the context of $R$-Parity violation, we find that the
indirect contributions are consistently around an order of magnitude or
more less than the direct ones, so that for this process and for the
similar ones of $B^0-\bar B^0$ and $D^0-\bar D^0$ mixing we cannot
improve on results in the literature \cite{bgnn,ag,crs,chr}.

\noindent {\bf Acknowledgments \hfil} \\
PLW would like to thank Jonathan Flynn for teaching him something about
QCD. BdC is very grateful to the University of Oxford for their
hospitality during the early stages of this work.
The work of BdC was supported by a PPARC Postdoctoral Fellowship.

\appendix
\section{Renormalisation Group Equations}
We now explain how to derive the RGEs which we shall use. Full RGEs for
our model are given in reference \cite{paper1}, but there are two
slight complications which arise when considering quark flavour
violation which do not occur in the lepton flavour violating processes
considered there. The first of these is that the MSSM (and the SM)
contain QFV explicitly through the CKM matrices and through the MSSM
soft trilinears. The second is that the RGEs of \cite{paper1} are
presented in a weak basis while we work in a fermion mass eigenstate
basis.

The inclusion of the CKM matrix is straightforward, since defining
\be
\Lambda^d_{ij}=\delta_{ij}h^d_{j}
\ee
with $h^d_i$ one of the eigenvalues of the Yukawa matrix $h^d_{ij}$, and
similarly for $\Lambda^u$, $\Lambda^e$, we can simply put
\be
\lp{4ij}=-h^d_{ij}=-K^*\Lambda^d=-\Lambda^d-k\Lambda^d
\ee
where $K$ is the CKM matrix, which we shall henceforth consider to be
real, defining $K=1+k$ so that $k$ is anti-symmetric to first order in
$k$. A full description of our other notation is contained in reference
\cite{paper1}. This leads to off-diagonal terms in the running of the
soft squark masses
\br
16\pi^2 {d m^2_{Q_iQ_j} \over dt}&\simeq& 
         k_{ij}({h_j^d}^2-{h_i^d}^2)(m^2_{Q_i}+m^2_{Q_j}+2m^2_{H_1})\cr
   &+& 2 k_{ij}({h_j^d}^2m^2_{d_j}-{h_i^d}^2m^2_{d_i})
       + 2 \eta^u_{ik}\eta^u_{jk}
       + 2 \eta^d_{ik}\eta^d_{jk} \cr
16\pi^2 {d m^2_{d_id_j} \over dt}&\simeq&
       4 k_{ij}h_i^dh_j^d(m^2_{Q_i}-m^2_{Q_j})
       + 4 \eta^d_{ki}\eta^d_{kj}
\label{CKM}
\er
which are in addition to the $R$-Parity violating contributions. Here
$\eta^d$ is given at the unification scale by
$\eta^d=h^dA_0=K^*\Lambda^dA_0$, but in general need not run in such a
way as to preserve this relationship, and we have assumed that the
diagonal terms in the mass matrix $m^2_x:=m^2_{xx}$ are very much
larger than the off-diagonal ones. Including the down and lepton soft
trilinears in the equations given in \cite{paper1} can be done by
treating $H_1$ as $L_4$ as noted there.

Including the full matrix of $\eta^u_{ij}$ is straightforward. We
give here the full RGE
\br
16\pi^2 {d \eta^u_{ij} \over dt}&=& 
   \eta^u_{ij} \bigl( 
     3{h_k^u}^2 + 5{h_i^u}^2 + 4{h_j^u}^2 \bigr )
           + 6\delta_{ij} h_i^uh_k^u\eta^u_{kk} \cr
       &&  + 4h_i^u\lpp{imn}\Cpp{jmn}
           + 2\eta^u_{ik}\lpp{kmn}\lpp{jmn}
           + \eta^u_{kj}\lp{mkn}\lp{min}
           + 2h_j^u\lp{mjn}\Cp{min}\cr
       &&  - \eta^u_{ij} \left (
               {13\over 9}g_1^2 + 3g_2^2 + {16\over 3}g_3^2\right ) \cr
       && + 2\delta_{ij} h_i^u \left ({13\over 9}g_1^2M_1 + 3g_2^2M_2
                       + {16\over 3}g_3^2M_3 \right )
\er
where we have included the down and lepton Yukawa couplings and
trilinear soft terms only implicitly ({\it i.e.} through the $\l{}$,
$\lp{}$, $\C{}$, and $\Cp{}$). We also must then include a term in the
RGE for $\Cp{ijk}$ of  $2\eta^u_{jl}h_l^u\lp{ilk}+{h_j^u}^2\Cp{ijk}$
and one in the RGE for $\Cpp{ijk}$ of
$4\eta^u_{li}h_l^u\lpp{ljk}+2{h_i^u}^2\Cpp{ijk}$. Inclusion of the
remaining Yukawa couplings is trivial.

The second and rather more complicated issue is that of the difference
between the weak and mass bases. We have three Yukawa matrices $h^u$,
$h^d$, $h^e$, and it is conventional to choose a basis such that
\be
h^u=\Lambda^u \qquad\qquad
h^d=K\Lambda^d \qquad\qquad
h^e=\Lambda^e
\label{basis}
\ee
where $K$ is now being taken real.
Such a choice is also a weak eigenstate basis. However, these relations
are not stable under the RGEs with either a non-trivial CKM matrix or
$R$-Parity violating couplings. Hence we will find in general that all of
the Yukawa matrices will have non-zero off-diagonal elements, greatly
complicating our analysis. The simplest way of dealing with this is to
define a basis where the field rotations to impose equation~(\ref{basis})
are performed in a scale-dependent way. Then equation~(\ref{basis}) will
always be satisfied, but the RGEs will differ slightly.

If we define
\be
\Lambda^u=Uh^uV_u^T
\ee
where $U$, $V$ are orthogonal matrices with explicit dependence on the
renormalisation scale, then
\be
\dot{\Lambda^u}=
\dot Uh^uV_u^T + U\dot h^uV_u^T + Uh^u\dot V_u^T
\ee
where the dots indicate $16\pi^2d/dt$ and defining $u=\dot UU^T$ and
$v^u=\dot V_uV_u^T$ we find that
\be
\dot{\Lambda^u}=
u\Lambda^u - \Lambda^uv^u + \gamma^Q\Lambda^u
+\Lambda^u(\gamma^u+\gamma^{H_2})
\ee
where the anomalous dimension matrices $\gamma$ are solved in the basis
where equation~(\ref{basis}) is satisfied. Hence we can solve for $u$
and $v^u$ in terms of the Yukawa couplings and anomalous dimensions.
The situation for the down-type Yukawas is slightly more complicated.
We choose to define
\be
\Lambda^d=K^T(Wh^dV_d^T)
\ee
where $K$ is the CKM matrix and $W$ and $V_d$ are unitary as before.
Hence we may calculate $w$ and $v^d$ defined by $w=\dot WW^T$ and
$v^d=\dot V_dV_d^T$. For the leptonic case we define $u^e$ and $v^e$
exactly as for up-type quarks.

Imposing that the $\Lambda$ remain diagonal we find
\br
u_{ij}&=& -\gamma^Q_{ij}\left (
  \frac{ {h_j^u}^2+{h_i^u}^2 }{ {h_j^u}^2-{h_i^u}^2 }\right )
          -\gamma^u_{ij}\left (
  \frac{ 2h_i^uh_j^u }{ {h_j^u}^2-{h_i^u}^2 }\right )\cr
v^u_{ij}&=& -\gamma^Q_{ij}\left (
  \frac{ 2h_i^uh_j^u }{ {h_j^u}^2-{h_i^u}^2 } \right )
          -\gamma^u_{ij}\left (
  \frac{ {h_j^u}^2+{h_i^u}^2 }{ {h_j^u}^2-{h_i^u}^2 } \right) \cr
\tilde w_{ij}&=& -\tilde\gamma^Q_{ij}\left (
  \frac{ {h_j^d}^2+{h_i^d}^2 }{ {h_j^d}^2-{h_i^d}^2 }\right )
          -\gamma^d_{ij}\left (
  \frac{ 2h_i^dh_j^d }{ {h_j^d}^2-{h_i^d}^2 }\right )\cr
v^d_{ij}&=& -\tilde\gamma^Q_{ij}\left (
  \frac{ 2h_i^dh_j^d }{ {h_j^d}^2-{h_i^d}^2 } \right )
          -\gamma^d_{ij}\left (
  \frac{ {h_j^d}^2+{h_i^d}^2 }{ {h_j^d}^2-{h_i^d}^2 } \right) \cr
u^e_{ij}&=& -\gamma^L_{ij}\left (
  \frac{ {h_j^e}^2+{h_i^e}^2 }{ {h_j^e}^2-{h_i^e}^2 }\right )
          -\gamma^e_{ij}\left (
  \frac{ 2h_i^eh_j^e }{ {h_j^e}^2-{h_i^e}^2 }\right )\cr
v^e_{ij}&=& -\gamma^L_{ij}\left (
  \frac{ 2h_i^eh_j^e }{ {h_j^e}^2-{h_i^e}^2 } \right )
          -\gamma^e_{ij}\left (
  \frac{ {h_j^e}^2+{h_i^e}^2 }{ {h_j^e}^2-{h_i^e}^2 } \right)
\er
Here we use $\tilde w=K^TwK$ and $\tilde\gamma=K^T\gamma K$.

Explicitly expanding taking the masses to be very strongly ordered with
$h_1\ll h_2\ll h_3$, we find that for $i>j$
\br
u_{ij}&=&
        - k_{ij} {h_i^d}^2 + \sum_{mn}\lp{min}\lp{mjn}
        + 4 \frac{h_j^u}{h_i^u}\sum_{mn}\lpp{imn}\lpp{jmn}\cr
v^u_{ij}&=&
        - 2\frac{h_j^u}{h_i^u}k_{ij} {h_i^d}^2
        + \sum_{mn}2\lpp{imn}\lpp{jmn}
        + 2 \frac{h_j^u}{h_i^u}\sum_{mn}\lp{min}\lp{mjn}\cr
w_{ij}&=&
          k_{ij} {h_i^u}^2 + \sum_{mn}\lp{min}\lp{mjn}
        + 4 \frac{h_j^d}{h_i^d}\sum_{mn}(\lp{mni}\lp{mnj}
                  + \lpp{mni}\lpp{mnj})\cr
v^d_{ij}&=&
        2 \frac{h_j^d}{h_i^d}k_{ij} {h_i^u}^2
        + 2\sum_{mn}(\lp{mni}\lp{mnj}+\lpp{mni}\lpp{mnj})
        + 2 \frac{h_j^d}{h_i^d}\sum_{mn}\lp{min}\lp{mjn}\cr
u^e_{ij}&=&
         \sum_{mn}(\l{imn}\l{jmn}+3\lp{imn}\lp{jmn})
        + 4 \frac{h_j^e}{h_i^e}\sum_{mn}\l{mni}\l{mnj}\cr
v^e_{ij}&=&
          2 \sum_{mn}\l{mni}\l{mnj}
        + 2 \frac{h_j^e}{h_i^e}
               \sum_{mn}(\l{imn}\l{jmn}+3\lp{imn}\lp{jmn})\cr
u_{ij}-w_{ij} &=&
          -k_{ij} ({h_i^u}^2+{h_i^d}^2)
        + 4 \frac{h_j^u}{h_i^u}\sum_{mn}\lpp{imn}\lpp{jmn}\cr
      &&\qquad\qquad\qquad - 4 \frac{h_j^d}{h_i^d}\sum_{mn}(\lp{mni}\lp{mnj}
                  + \lpp{mni}\lpp{mnj})
\er

Hence we can simply use the RGEs in reference \cite{paper1}, but adding
the explicit CKM contributions and including
the conversion of basis by, for example
\be
16\pi^2 {d m^2_{Q_iQ_j} \over dt}^{mass\ basis}=
16\pi^2 {d m^2_{Q_iQ_j} \over dt}^{weak\ basis}
 +\sum_k ( u_{ik}m^2_{Q_kQ_j} +u_{jk}m^2_{Q_iQ_k} )
\ee

This gives the fermion mass basis RGEs for the off-diagonal terms in
the soft masses
\br
16\pi^2 {d m^2_{Q_iQ_j} \over dt}&=& 
         \sum_{mn} \Bigl(\lp{min}\lp{mjn}
                  \bigl(m^2_{Q_i} + m^2_{Q_j}
                     + 2m^2_{d_n} + 2m^2_{L_m}\bigr)
                     + 2\Cp{min}\Cp{mjn}\Bigr) \cr
           && + k_{ij}({h_j^d}^2-{h_i^d}^2)(m^2_{Q_i}+m^2_{Q_j})
             +2 k_{ij}({h_j^d}^2m^2_{d_j}-{h_i^d}^2m^2_{d_i}) \cr
           && \qquad\qquad
                  + 2 k_{ij}({h_j^d}^2-{h_i^d}^2)m^2_{H_1} \cr
           &&     + 2 \eta^u_{ik}\eta^u_{jk}
                  + 2 \eta^d_{ik}\eta^d_{jk}
                  + u_{ij}(m^2_{Q_j}-m^2_{Q_i})\\
16\pi^2 {d m^2_{u_iu_j} \over dt}&=& 
         \sum_{mn} \Bigl(\lpp{imn}\lpp{jmn}
                 \bigl(m^2_{u_i} + m^2_{u_j} + 4m^2_{d_n}\bigr)
               + 2\Cpp{imn}\Cpp{jmn}\Bigr) \cr
            && \qquad
                  + 4 \eta^u_{ki}\eta^u_{kj}
                  + v^u_{ij}(m^2_{u_i}-m^2_{u_j})\\
16\pi^2 {d m^2_{d_id_j} \over dt}&=&
         \sum_{mn}\Bigl(2\lp{mni}\lp{mnj}
                 \bigl(m^2_{d_i} + m^2_{d_j}
                    + 2m^2_{Q_n} + 2m^2_{L_m}\bigr)
                    + 4\Cp{mni}\Cp{mnj} \Bigr)\cr
      &+&\sum_{mn}\Bigl(2\lpp{mni}\lpp{mnj}
                 \bigl(m^2_{d_i}+m^2_{d_j}
                    + 2m^2_{d_n} + 2m^2_{u_m}\bigr)
                    + 4\Cpp{mni}\Cpp{mnj}\Bigr) \cr
            && + 4 k_{ij}h_i^dh_j^d(m^2_{Q_i}-m^2_{Q_j})
                    + 4 \eta^d_{ki}\eta^d_{kj}
                    + v^d_{ij}(m^2_{d_i}-m^2_{d_j})\\
16\pi^2 \frac{d m^2_{L_iL_j}}{dt} & = &
         \sum_{mn} \Bigl(\l{imn}\l{jmn}
                  \bigl(m^2_{L_i} + m^2_{L_j}
                     + 2m^2_{e_n} + 2m^2_{L_m}\bigr)
                     + 2\C{imn}\C{jmn} \Bigr ) \cr
      &+& \sum_{mn}\Bigl(3\lp{imn}\lp{jmn}
                  \bigl(m^2_{L_i} + m^2_{L_j}
                     + 2m^2_{d_n} + 2m^2_{Q_m}\bigr)
                     + 6\Cp{imn}\Cp{jmn}\Bigr) \cr
            && \qquad
                     + 2 \eta^e_{ik}\eta^e_{jk}
                     + u^e_{ij}(m^2_{L_j}-m^2_{L_i})\\
16\pi^2 \frac{d m^2_{e_ie_j}}{dt} & = &
         \sum_{mn} \Bigl(\l{mni}\l{mnj}
             \bigl(m^2_{e_i} + m^2_{e_j}
                + 4m^2_{L_n} \bigr)
                + 2\C{mni}\C{mnj} \Bigr) \cr
           && \qquad
                + 4 \eta^e_{ki}\eta^e_{kj}
                + v^e_{ij}(m^2_{e_i}-m^2_{e_j})
\er
It is clear from these equations that when we consider the first and
second generations, where the soft masses are nearly degenerate, the
difference in basis is totally irrelevant, but for cases involving the
third generation which is typically rather lighter the effects may be
non-negligible.

The RGE for the CKM matrix is then
\be
16\pi^2 \frac{d k_{ij}}{dt} = u_{ij}-w_{ij}
\ee

\section{Approximate Solutions of RGEs}
For our discussion of the indirect contributions to the various
processes, it will now be helpful to present some approximate formulae
for the various off-diagonal mass insertions. For convenience we will
present here formulae for all the elements, not merely those involved
in $b\to s\gamma$. If we begin by switching off the CKM matrix we find
\begin{eqnarray}
m^2_{Q_iQ_j}(M_Z) & = & - \sum_{mn} \lp{min}\lp{mjn}(M_{GUT})
  (5m_0^2 + 15M_{1/2}^2 - 5 A_0M_{1/2} + A_0^2 ) \nonumber \\
m^2_{d_id_j}(M_Z) & = & - \sum_{mn} \lp{mni}\lp{mnj}(M_{GUT})
  (10m_0^2 + 30M_{1/2}^2 - 10 A_0M_{1/2} +2 A_0^2 ) \nonumber \\
   && - \sum_{mn} \lpp{mni}\lpp{mnj}(M_{GUT})
  (10m_0^2 + 50M_{1/2}^2 - 20 A_0M_{1/2} + 5 A_0^2 ) \nonumber \\
m^2_{u_iu_j}(M_Z) & = & - \sum_{mn} \lpp{imn}\lpp{jmn}(M_{GUT})
  (10m_0^2 + 50M_{1/2}^2 - 20 A_0M_{1/2} + 5 A_0^2 ) \nonumber \\
\label{dmapp}
\end{eqnarray}
while for the $\Delta m^2_{LR}$ and $\Delta m^2_{RL}$ insertions we find
\begin{eqnarray}
\eta_{12}^d\nu_1 & = & m_s \sum_{mn} \lp{m1n}\lp{m2n}(M_{GUT})
  (3M_{1/2} - 1.5 A_0) \nonumber \\
\eta_{21}^d\nu_1 & = & m_s \sum_{mn} \lp{mn1}\lp{mn2}(M_{GUT})
  (9M_{1/2} - 4.5 A_0) \nonumber \\
               & + & m_s \sum_{mn} \lpp{mn1}\lpp{mn2}(M_{GUT})
  (9M_{1/2} - 3.5 A_0)
\label{dmLRapp}
\end{eqnarray}
where $m_s$ is the running strange mass at $M_Z$. Since all these
formulae depend quite strongly on $\alpha_3$ they should not be trusted
to better than about a factor of two. Note that the
$\Delta m^2_{LR}$ and $\Delta m^2_{RL}$ insertions are very much smaller
than the others because they are proportional to a soft mass times a
strange quark mass rather than two soft masses.

Including the CKM matrix is straightforward, and leads extra
contributions to those given above of
\begin{eqnarray}
m^2_{Q_1Q_2}(M_Z) & = & -\frac{10^{-7}}{\cos^2\beta}
  (3m_0^2 + 10M_{1/2}^2 - 4 A_0M_{1/2} + A_0^2 ) \nonumber \\
m^2_{Q_1Q_3}(M_Z) & = & -\frac{10^{-6}}{\cos^2\beta}
  (2m_0^2 + 8M_{1/2}^2 - 3 A_0M_{1/2} + 0.8A_0^2 ) \nonumber \\
m^2_{Q_2Q_3}(M_Z) & = & -\frac{10^{-5}}{\cos^2\beta}
  (2m_0^2 + 8M_{1/2}^2 - 3 A_0M_{1/2} + 0.8A_0^2 ) \nonumber \\
\eta_{12}^d\nu_1 & = & m_s (-0.8M_{1/2} + 0.2 A_0) \cr
\eta_{21}^d\nu_1 & \ll & \eta_{12}^d\nu_1
\label{dmappCKM}
\end{eqnarray}
although here the errors are larger, owing to the dependence on poorly
known Yukawa couplings. Although we have only shown the contributions to
$m^2_{Q_iQ_j}$, in fact there are also contributions
to $m^2_{d_2d_3}$ and $m^2_{u_2u_3}$ in particular which are much
smaller but non-zero.

\section{Definition of Functions}
The expressions for the amplitudes in $\bsg$ and $K^0\bar K^0$ mixing 
presented earlier employ a number of functions, many of which are
contained in \cite{paper1}, and the remainder of which are given
explicitly here. Functions occurring in $b\to s\gamma$ are
\br
f_{\gamma}^{(1)}(x) &=& Q_u F_1(x)+F_2(x)\cr 
f_g^{(1)}&=&F_1(x)
\cr
f_{\gamma}^{(2)}(x) &=& Q_u F_3(x)+F_4(x) \cr
f_g^{(2)}&=&F_3(x)
\cr
f_{\gamma}^{(3)}(x) &=& F_1(x) + Q_u F_2(x) \cr
f_g^{(3)}&=&F_2(x)
\cr
f_{\gamma}^{(4)}(x) &=& F_3(x) + Q_u F_4(x) \cr
f_g^{(4)}&=&F_4(x)
\cr
f_{\gamma}^{(5)}(x) &=& -Q_u F_1(x)+Q_d F_2(x) \cr
f_g^{(5)}&=&\frac{1}{2}F_1(x)-\frac{1}{2}F_2(x)
\cr
f_{\gamma}^{(6)}(x) &=& -Q_d F_1(x)+Q_u F_2(x) \cr
f_g^{(6)}(x) &=& f_g^{(5)}(x)
\er

In addition, for the $\Delta m_K$ calculation 
the functions $I_4$ and $J_4$ used in the text are given by
\br
I_4(m_1^2,m_2^2,m_3^2,m_4^2)&=&
\int_0^1dxdydz \frac{z(1-z)}{D_4} \cr
J_4(m_1^2,m_2^2,m_3^2,m_4^2)&=&
\int_0^1dxdydz \frac{z(1-z)}{D_4^2}
\er
where we have defined
\be
D_4(m_1^2,m_2^2,m_3^2,m_4^2)
=[xm_1^2+(1-x)m_2^2]z + [ym_3^2+(1-y)m_4^2](1-z)
\ee
Note that $I_4$ and $J_4$ are totally symmetric about interchange of
arguments, and can be simplified in certain limits as follows.
\br
I_4(m^2,m^2,M^2,M^2)&=& 
  \frac{1}{M^2}F_4(x) \\
I_4(m_1^2,m_2^2,M^2,0)&=&
  \frac{1}{M^2}\Biggl [
  \frac{ (1-x_1)x_2\ln x_2 - (1-x_2) x_1\ln x_1}
       { 2(x_1-1)(x_1-x_2)(x_2-1)} 
\Biggr ]\\
I_4(m_1^2,m_2^2,M^2,M^2)&=&
  \frac{1}{M^2}\Biggl [
   \frac{1}{2(1-x_1)(1-x_2)} \cr
&& \qquad\qquad
 + \frac{1}{x_2-x_1}\Biggl (
         \frac{x_2^2\ln(x_2)}{2(1-x_2)^2}
       - \frac{x_1^2\ln(x_1)}{2(1-x_1)^2}
      \Biggr ) \Biggr ] \\
I_4(m^2,M^2,0,0)&=&
  \frac{1}{M^2} \Biggl [\frac{\ln x}{2(x-1)} \Biggr ]\\
I_4(m^2,m^2,M^2,0)&=& 
  \frac{1}{2m^2}\Biggl [\frac{1-y+y\ln(y)}{(1-y)^2} \Biggr ]\\
I_4^{\prime\prime}(m^2,m^2,M^2,M^2)&=&
  \frac{1}{m^6}G(y) = - \frac{1}{2m^6} \tilde f_6(x) \\
I_4^{\prime\prime}(m^2,m^2,M_1^2,M_2^2)&=& 
  \frac{1}{m^4(M_1^2-M_2^2)}\Biggl [
       y_1F_1(y_1)-y_2F_1(y_2)   \Biggr ] \\
I_4^{\prime\prime}(m_1^2,m_2^2,M^2,M^2)&=& 
  \frac{1}{M^6}\Biggl [
   \frac{x_1+x_2+x_1^2+x_2^2-6x_1x_2+x_1^2x_2+x_1x_2^2}
        {2(x_1-1)^2(x_2-1)^2(x_1-x_2)^2} \cr
&&
       + \frac{(x_1x_2-x_1^3)\ln(x_1)}{(x_1-1)^3(x_1-x_2)^3}
       - \frac{(x_1x_2-x_2^3)\ln(x_2)}{(x_2-1)^3(x_1-x_2)^3}
\Biggr ] \\
J_4(m_1^2,m_2^2,M^2,M^2)&=&
  \frac{1}{M^4}\Biggl [
   \frac{-1}{(1-x_1)(1-x_2)} \cr
&& \qquad\qquad
 + \frac{1}{x_2-x_1}\Biggl (
         \frac{x_1\ln(x_1)}{(1-x_1)^2}
       - \frac{x_2\ln(x_2)}{(1-x_2)^2}
      \Biggr ) \Biggr ] \\
J_4(0,m_1^2,m_2^2,M^2)&=&
  \frac{1}{M^4}\Biggl [
   \frac{(x_2-1)\ln(x_1)-(x_1-1)\ln(x_2)}
        {(x_2-1) (x_2-x_1) (x_1-1)} \Biggr ]
\\
J_4^{\prime\prime}(m^2,m^2,M^2,M^2)&=& 
  \frac{2}{m^4M^4}F(x) = \frac{1}{m^8}xf_6(x)\\
J_4^{\prime\prime}(m_1^2,m_2^2,M^2,M^2)&=& 
  \frac{1}{M^8}\Biggl [ 
    \frac{-2(1-x_1-x_2+x_1^2+x_2^2-x_1x_2)}
         { (x_1-1)^2 (x_2-1)^2 (x_1-x_2)^2} \cr
&& \qquad
 - \frac{(x_1+x_2+x_1x_2-3x_1^2)\ln(x_1)} { (x_1-x_2)^3 (x_1-1)^3 }
\cr
&& \qquad
 + \frac{(x_1+x_2+x_1x_2-3x_2^2)\ln(x_2)} { (x_1-x_2)^3 (x_2-1)^3 }
 \Biggr ]
\\
J_4^{\prime\prime}(m^2,m^2,M_1^2,M_2^2)&=&
  \frac{1}{m^8}\Biggl [ 
   \frac{y_2\ln(y_2)} {(y_2-1)^4(y_1-y_2)}
 - \frac{y_1\ln(y_1)} {(y_1-1)^4(y_1-y_2)} \cr
&& + \frac{1}
{6(y_1-1)^3(y_2-1)^3}
 \biggl (
 -11 + 7(y_1+y_2) \\
&& \qquad
- 2 (y_1^2+y_2^2-5y_1y_2) -5(y_1+y_2)y_1y_2+ y_1^2y_2^2 \biggr )
\Biggr ] \nonumber
\er
Here $x=m^2/M^2$, $x_i=m_i^2/M^2$, $y=M^2/m^2$ and $y_i=M_i^2/m^2$, and
\br
I_4^{\prime\prime}(m_1^2,m_2^2,m_3^2,m_4^2)&=&
\frac{\partial^2}{\partial m_1^2 \partial m_2^2}
  I_4(m_1^2,m_2^2,m_3^2,m_4^2)\cr
J_4^{\prime\prime}(m_1^2,m_2^2,m_3^2,m_4^2)&=&
\frac{\partial^2}{\partial m_1^2 \partial m_2^2}
  J_4(m_1^2,m_2^2,m_3^2,m_4^2)
\er
We give $f_6$ and $\tilde f_6$ defined in reference~\cite{hagelin} for
convenience when comparing our results with other authors.

\newpage

\section{Figure Captions}
 
\begin{figure}[htb]
\epsfxsize=15cm
\centerline{\epsfbox{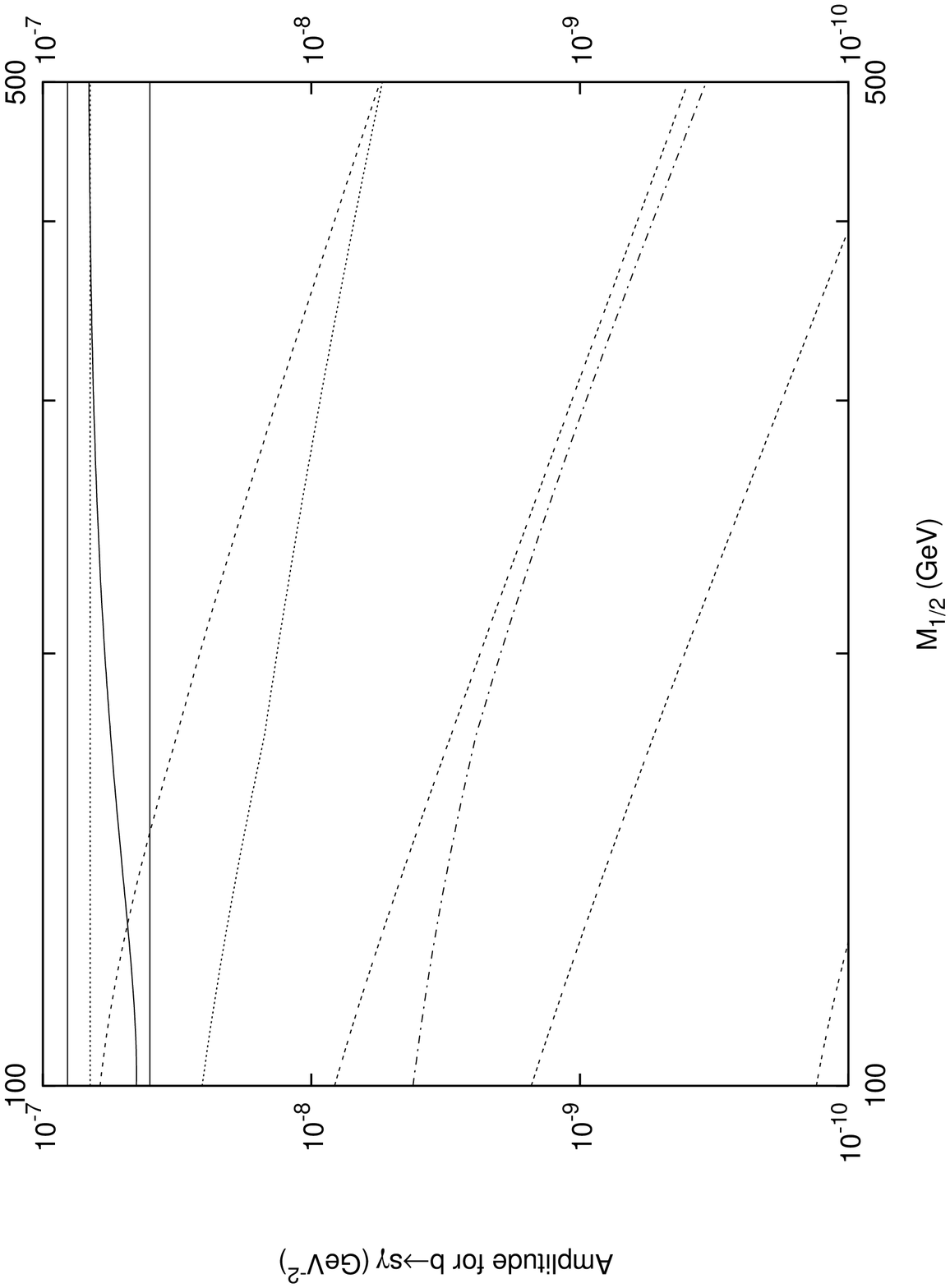}}
\caption{}
\label{bsglpLL}
\end{figure}

\noindent
{\bf Figure~\ref{bsglpLL}}\\
Absolute values for various contributions to the $b\to
s\gamma$ amplitude $\tilde A_{LR}$. Parameters are
$\lp{121}(M_{GUT})=\lp{131}(M_{GUT})=0.05$, $m_0=100$ GeV, $A_0=0$,
$\tan\beta=10$, $\mu_4<0$. Contributions from Higgs (dotted), SM
(horizontal dotted), neutralino (double dashes), chargino (dashes for
the MSSM part with triple dashes for the remainder), gluino (quadruple
dashes), direct RPV diagrams (dot dashed) and total (solid) are shown.
The experimental upper and lower limits are shown as horizontal solid
lines.

\begin{figure}[htb]
\epsfxsize=15cm
\centerline{\epsfbox{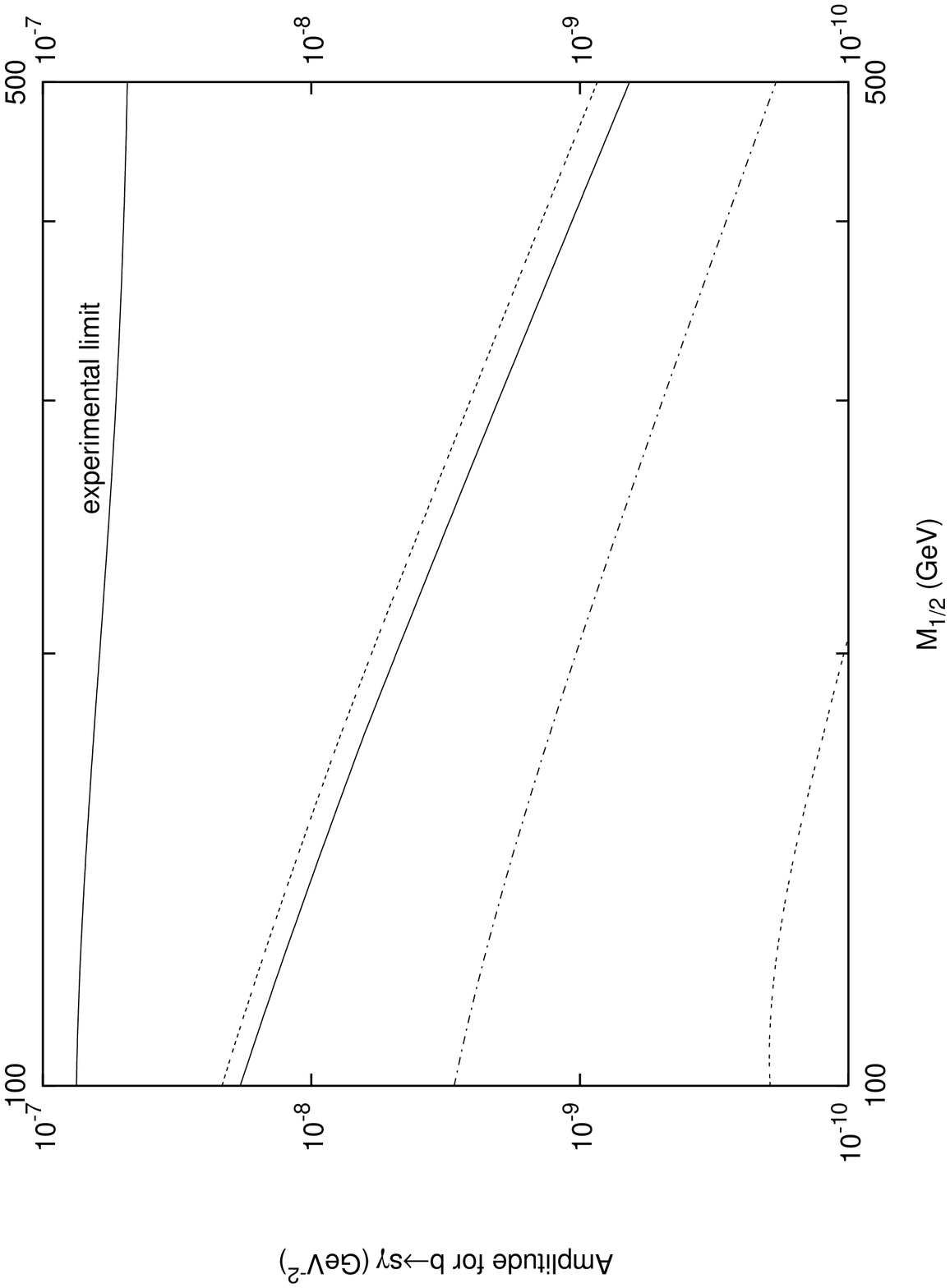}}
\caption{}
\label{bsglppRR}
\end{figure}

\noindent
{\bf Figure~\ref{bsglppRR}}\\
Absolute values for various contributions to the $b\to s\gamma$ amplitude
$\tilde A_{RL}$. Parameters are
$\lpp{112}(M_{GUT})=\lpp{113}(M_{GUT})=0.1$, $m_0=100$ GeV, $A_0=0$,
$\tan\beta=10$, $\mu_4<0$. Contributions from neutralino (double 
dashes), gluino (quadruple dashes), direct RPV diagrams (dot dashed) 
and total (solid) are shown.
The experimental upper limit shown varies with $M_{1/2}$ because the
MSSM part of $\tilde A_{LR}$ does.

\begin{figure}[htb]
\epsfxsize=15cm
\centerline{\epsfbox{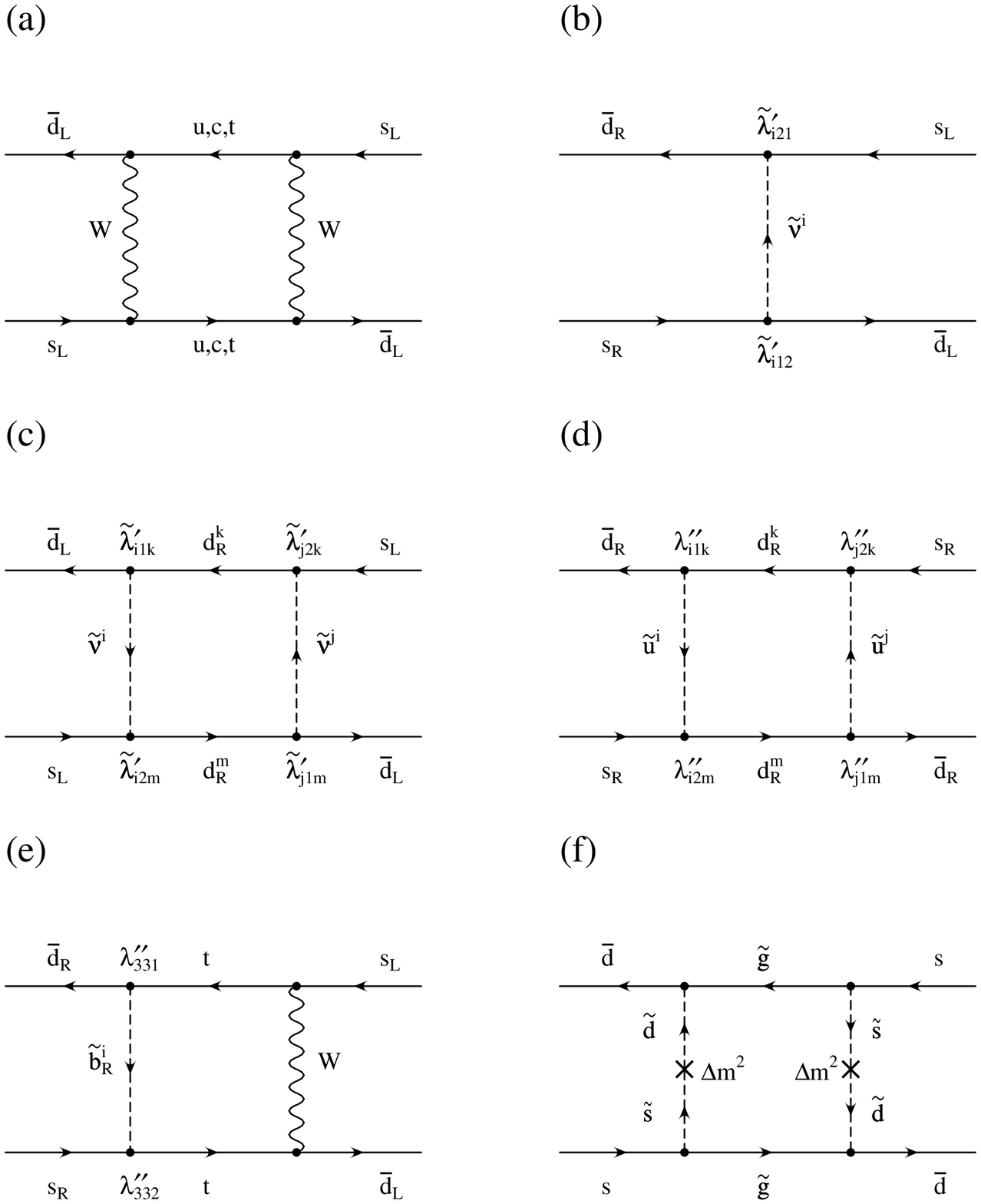}}
\caption{}
\label{KKdiag}
\end{figure}

\noindent
{\bf Figure~\ref{KKdiag}}\\
Some of the diagrams contributing to $K^0-\bar K^0$ mixing.

\begin{figure}[htb]
\epsfxsize=15cm
\centerline{\epsfbox{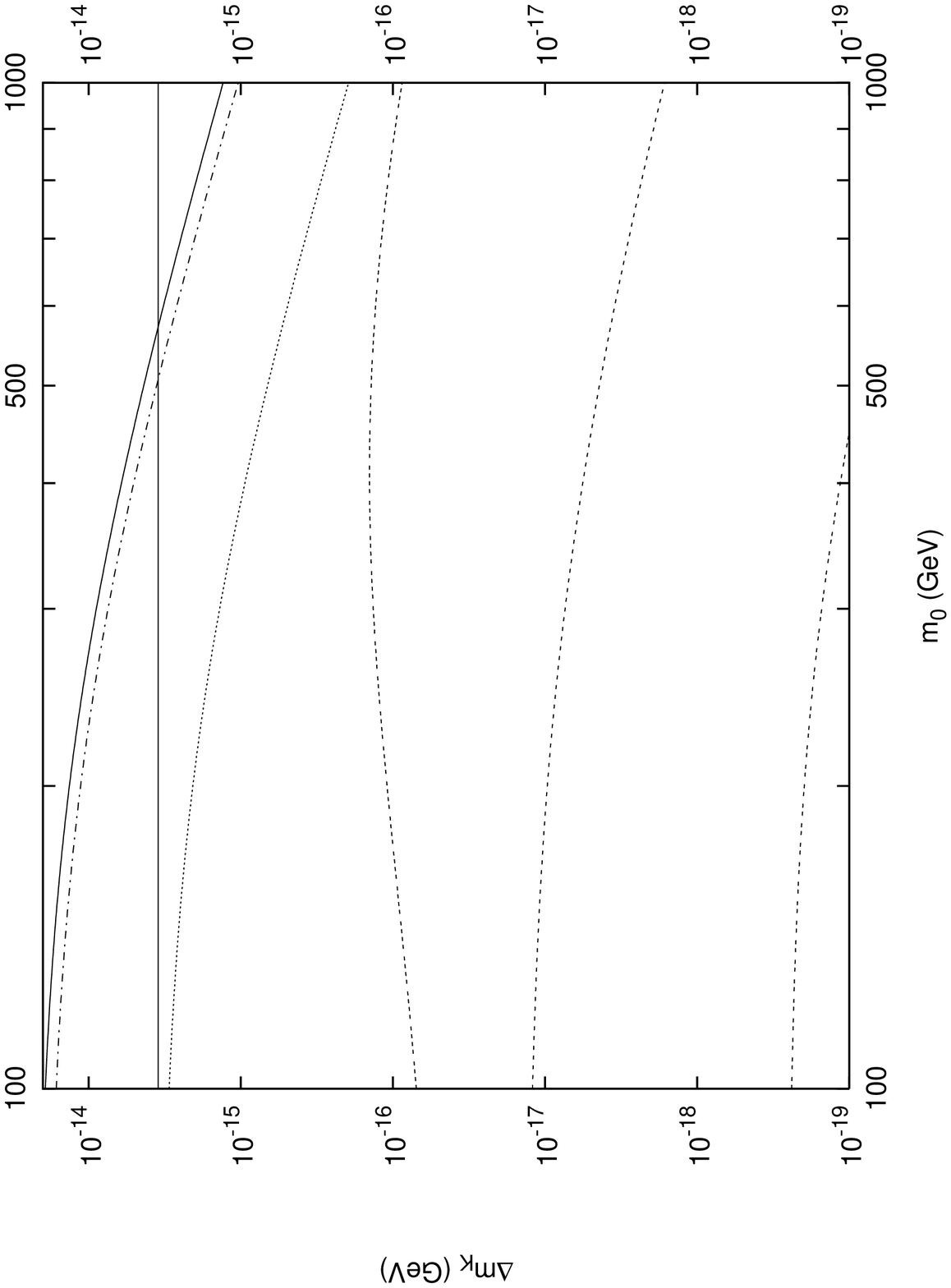}}
\caption{}
\label{KKbarlppRR}
\end{figure}

\noindent
{\bf Figure~\ref{KKbarlppRR}}\\
Absolute values of various contributions to $\Delta m_K$. Parameters
are $\lpp{213}(M_{GUT})=\lpp{223}(M_{GUT})=0.02$, $M_{1/2}=100$ GeV,
$A_0=0$, $\tan\beta=10$, $\mu_4<0$. Contributions from neutralino
(double dashes), gluino (quadruple dashes), mixed gluino-neutralino
(dashed), direct RPV diagrams with and without $W$ lines (dotted and
dot dashed respectively) and total (solid) are shown, together with the
experimental upper limit (horizontal solid line).

\begin{thebibliography}{99}
\bibitem{revs}
For reviews see for example 
H.P.~Nilles, {\it Phys. Rep.} {\bf 110} (1984) 1;\\
H.E.~Haber and G.L.~Kane, {\it Phys. Rep.} {\bf 117} (1985) 75.
%
\bibitem{rpv}
C.S.~Aulah, R.N.~Mohapatra, {\it Phys. Lett.} {\bf B119} (1982) 316; \\
F.~Zwirner, {\it Phys. Lett.} {\bf B132} (1983) 103;\\
S.~Dawson, {\it Nucl. Phys.} {\bf B261} (1985) 297;\\
S.~Dimopoulos, L.J.~Hall, {\it Phys. Lett.} {\bf B196} (1987) 135.
%
\bibitem{barbm}
R.~Barbieri, A. Masiero, {\it Nucl. Phys.} {\bf B267} (1986) 679.
%
\bibitem{suzuki}
L.J.~Hall, M.~Suzuki, {\it Nucl. Phys.} {\bf B231} (1984) 419.
%
\bibitem{bhat} For a summary of experimental limits on $R$-Parity
violating couplings, see G.~Bhattacharyya, hep-ph/9608415.
%
\bibitem{bgh} V.~Barger, G.F.~Giudice, T.~Han,
 {\it Phys. Rev.} {\bf D40} (1989) 2987.  
%
\bibitem{bpw} V.~Barger, R.J.N.~Phillips, K.~Whisnant,
 {\it Phys. Rev.} {\bf D44} (1991) 1629.  
%
\bibitem{bgnn} T.~Banks, Y.~Grossman, E.~Nardi, Y.~Nir,
 {\it Phys. Rev.} {\bf D52} (1995) 5319.
%
\bibitem{rv}
J.C.~Romao, J.W.F.~Valle, {\it Nucl. Phys.} {\bf B381} (1992) 87.
%
\bibitem{paper1} B.~de~Carlos, P.L.~White,
 {\it Phys. Rev.} {\bf D54} (1996) 3427.
%
\bibitem{hagelin}
J.~Hagelin, S.~Kelley, T.~Tanaka, {\it Nucl. Phys.} {\bf B415} (1994) 293; \\
J.~Hagelin, S.~Kelley, T.~Tanaka,
 {\it Mod. Phys. Lett.} {\bf A8} (1993) 2737.
%
\bibitem{otherfcnc}
L.J.~Hall, V.A.~Kostelecky, S.~Raby,
 {\it Nucl. Phys.} {\bf B267} (1986) 415; \\
F.~Gabbiani, A.~Masiero, {\it Nucl. Phys.} {\bf B322} (1989) 235; \\
G.C.~Branco, G.C.~Cho, Y.~Kizukuri, N.~Oshimo,
 {\it Phys. Lett.} {\bf B337} (1994) 316;\\
R.~Barbieri, L.~Hall, A.~Strumia,
 {\it Nucl. Phys.} {\bf B445} (1995) 219;
 {\it Nucl. Phys.} {\bf B449} (1995) 437; \\
E.~Gabrielli, A.~Masiero, L.~Silvestrini, 
 {\it Phys. Lett.} {\bf B374} (1996) 80; \\
T.~Goto, T.~Nihei, Y.~Okada,
 {\it Phys. Rev.} {\bf D53} (1996) 5233; \\
E.~Gabbiani, E.~Gabrielli, A.~Masiero, L.~Silvestrini,
 hep-ph/9604387.
%
\bibitem{rel}
A. Smirnov and F. Vissani, {\it Nucl. Phys.} {\bf B460} (1996) 37; \\
K. Tamvakis, hep-ph/9604343; \\
M. Bastero-Gil, B. Brahmachari, hep-ph/9606418; \\
D. Chang, W.Y. Keung, hep-ph/9608313.
%
\bibitem{pdb}
Particle Data Group Review of Particle Properties, 
 {\it Phys. Rev.} {\bf D54} (1996) 1.
%
\bibitem{CLEO} The CLEO Collaboration,
 {\it Phys. Rev. Lett.} {\bf 74} (1995) 2885.
%
\bibitem{bsgsusy}
See B.~de~Carlos, J.A.~Casas, Phys. Lett. {\bf B349} (1995) 300, 
and erratum in Phys. Lett. {\bf B351} (1995) 604, and references therein.
%
\bibitem{bsgQCD}
A.J. Buras, M. Misiak, M. M\"unz and S. Pokorski,
{\it Nucl. Phys.} {\bf B424} (1994) 374 and references therein; \\
K. Adel, Y.P. Yao, {\it Phys. Rev.} {\bf D49} (1994) 4945; \\
M. Ciuchini, E. Franco, G. Martinelli, L.Reina, L. Silvestrini, {\it
Phys. Lett.} {\bf B334} (1994) 137; \\
A. Ali, C. Greub, {\it Phys. Lett.} {\bf B361} (1995) 146; \\
N. Pott, {\it Phys. Rev.} {\bf D54} (1996) 938; \\
C. Greub, T. Hurth, D. Wyler, {\it Phys. Rev.} {\bf D54} (1996) 3350
and {\it Phys. Lett.} {\bf B380} (1996) 385.
%
\bibitem{bert} S.~Bertolini, F.~Borzumati, A.~Masiero, G.~Ridolfi,
 {\it Nucl. Phys.} {\bf B353} (1991) 591.
%
\bibitem{ag} K.~Agashe, M.~Graesser, {\it Phys. Rev.} {\bf D54} (1996)
4445.
%
\bibitem{gs}
J.L.~Goity, M.~Sher, {\it Phys. Lett.} {\bf B346} (1995) 69.
%
\bibitem{crs} C.E.~Carlson, P.~Roy, M.~Sher,
 {\it Phys. Lett.} {\bf B357} (1995) 99.
%
\bibitem{chr} D.~Choudhury, P.~Roy,
 {\it Phys. Lett.} {\bf B378} (1996) 153.
%
\end{thebibliography}
\end{document}